\documentclass[preprint,pteplogo]{ptephy_v2}

\preprintnumber{HK-TN-0121} 

\usepackage[
  colorlinks=true, 
  allcolors=blue,  
  breaklinks=true, 
  pdfencoding=auto
]{hyperref}
\usepackage{xurl} 

\usepackage{makecell}
\usepackage{array}

\usepackage{graphicx}

\begin{document}

\title{Excavation of a 69-m diameter and 94-m high cavern for the Hyper-Kamiokande detector}

\author{Y.~Asaoka$^{*}$}
\author{H.~Tanaka}
\author{S.~Nakayama}
\author{K.~Abe}
\author{K.~Ishita}
\author{S.~Moriyama}
\author{M.~Shiozawa}
\affil{Kamioka Observatory, Institute for Cosmic Ray Research, 
The University of Tokyo, 456 Higashi-Mozumi, Kamioka-cho, Hida-shi, Gifu, 506-1205 Japan \email{asaoka@km.icrr.u-tokyo.ac.jp}} 

\author{K.~Horinokuchi}
\affil{Mitsui Mineral Development Engineering Co., Ltd.,
1-11-1, Osaki, Shinagawa-ku, Tokyo 141-0032, Japan}

\author{C.~Miura}
\author{Y.~Suzuki}
\author{H.~Morioka}
\author{D.~Inagaki}
\author{H.~Kurose}
\author{T.~Suido}
\affil{Tokyo Electric Power Service CO., Ltd., 
9F, KDX Toyosu Grandsquare 7-12, 
Shinonome 1-chome, Koto-ku, Tokyo 135-0062, Japan}

\author{T.~Kobuchi}
\author{M.~Tobita}
\author{M.~Utsuno}
\affil{KAJIMA CORPORATION, 3-1, Motoakasaka 1-chome, 
Minato-ku, Tokyo 107-8388, Japan}

\begin{abstract}
The excavation of the Hyper-Kamiokande cavern, 600~m underground, is complete. Measuring 69~m in diameter and 94~m in height, it is among the world's largest rock caverns. A vertically oriented, dome-capped cylindrical design was chosen to optimize cost and performance. 
The geometry, combined with substantial overburden, posed major engineering challenges. 
This paper outlines the underground works, main cavern design, excavation plan, and the evolution of support design and construction methods during excavation, namely the information-based (observational) design and construction approach. 
\end{abstract}

\subjectindex{H20, H50}

\maketitle

\section{Introduction}
\begin{figure}[b!]
\begin{center}
\begin{minipage}[t]{0.345\linewidth}
\includegraphics[width=1.1\linewidth]{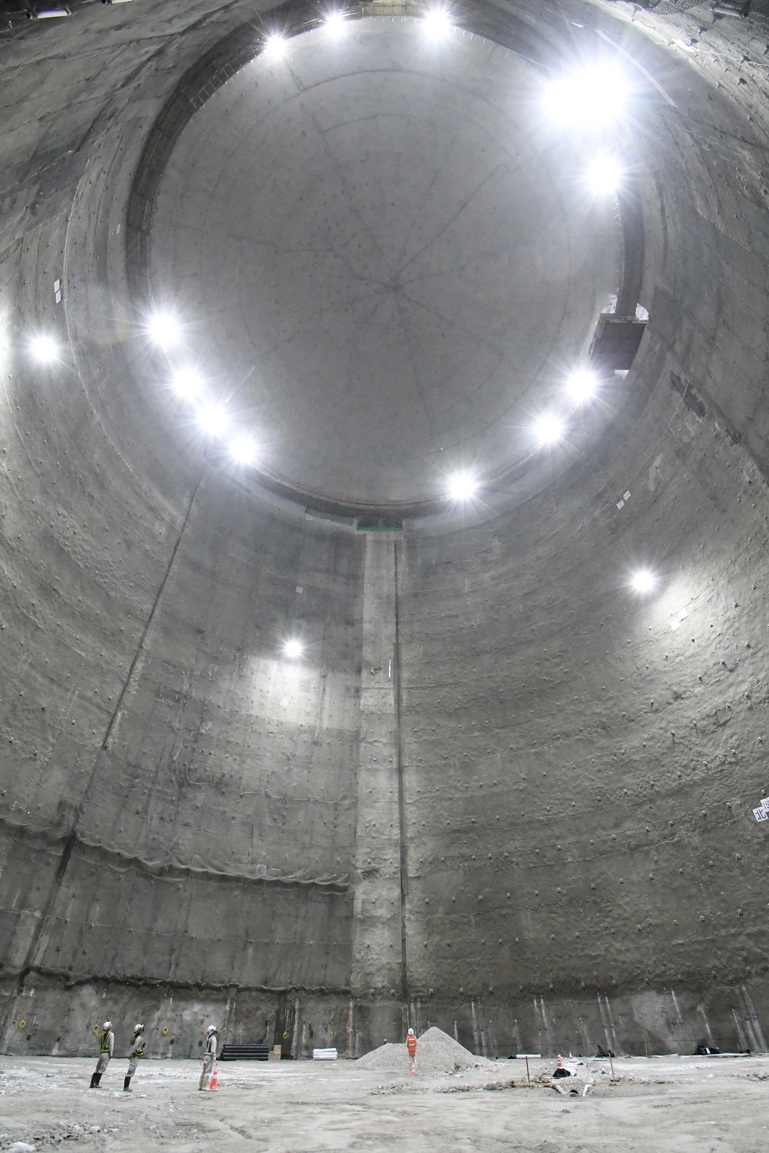}
\end{minipage}
\hspace{0.03\linewidth}
\begin{minipage}[t]{0.525\linewidth}
\includegraphics[width=1.1\linewidth]{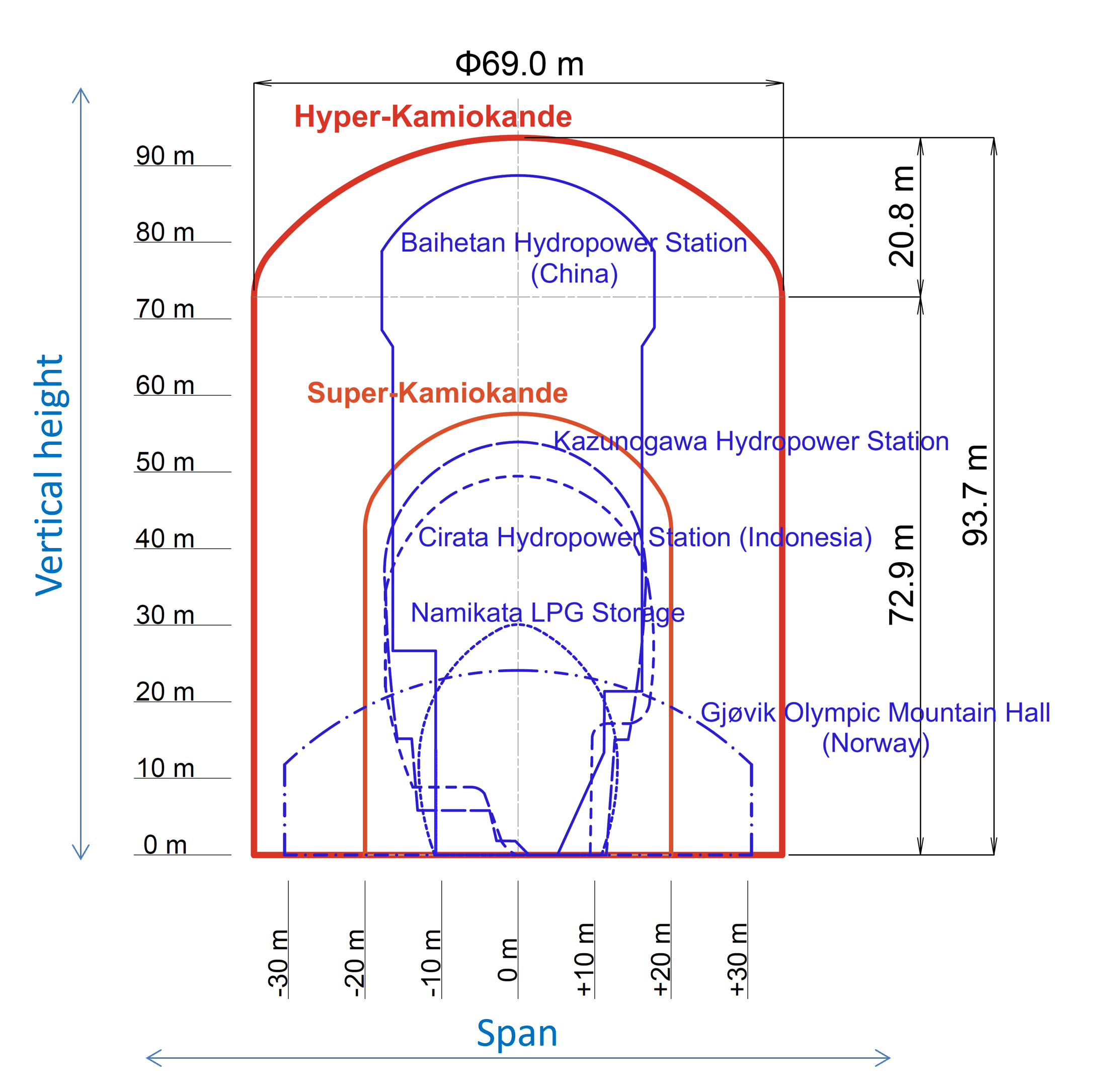} 
\hspace*{-1cm}
\vspace*{-0.5cm}
\end{minipage}
\end{center}
\vspace*{-0.5cm}
\caption{\small
(Left) Wide-angle photograph of the completed HK main cavern, showing the entire dome from the bottom of the cylindrical section. The HK cavern is located approximately 600~m underground in the mountains of Kamioka Town, Hida City, Gifu Prefecture, Japan. It has a diameter of 69~m, a height of 94~m, and a total volume of approximately 320,000~m$^3$. 
(Right) 
Span-versus-height cross-sectional comparison
of major underground rock caverns in Japan and abroad. 
Here, span denotes the shortest horizontal clear roof dimension across the excavated opening, a primary geometric
parameter controlling stability under in-situ stresses,
against gravity-driven block instability. 
Note: This span-versus-height representation may highlight vertically oriented caverns; the comparison is limited to the caverns listed in Table~\ref{tab:cavern_comp} (not exhaustive).
}
\label{fig:cavern_comp}
\end{figure}

Large underground caverns of exceptional span and depth pose engineering challenges not typically encountered in conventional underground-engineering practice. The Hyper-Kamiokande (HK) cavern, completed on 31 July 2025~\cite{HK_cavern}, represents such a case with its 69~m diameter (equal to the span in this case), 94~m height, and approximately 600~m of overburden. 
As the third-generation successor to Kamiokande~\cite{Kamiokande} and Super-Kamiokande (SK)~\cite{SK}, HK was designed to achieve an effective mass increase at the order-of-magnitude scale, while also satisfying substantial overburden requirements~\cite{HK-Koshiba,HK-Shiozawa,HK-Nakamura,HK-Itow,HK-Design}, driving the need for a cavern of unprecedented size. 
A photograph of the completed cavern together with a span-based comparison of major underground rock caverns worldwide is shown in Fig.~\ref{fig:cavern_comp}.

To address these engineering challenges, the project adopted an end-to-end integration of full 3D sequential elasto-plastic analyses with a prespecified support design-and-control framework, continuously updated using field measurements and geological observations---an information-based (observational) design and construction approach. This integration enabled timely, risk-informed decisions that ensured safety while optimizing cost and schedule under evolving ground conditions. Accordingly, we emphasize traceability from model assumptions to observations to design updates, so that the rationale behind key decisions is transparent and transferable. In this paper, we document this integrated design--construction process and its outcomes for the Hyper-Kamiokande cavern.

\begin{table}[bt!]
\caption{Comparison of characteristic sizes of man-made underground caverns worldwide.
Here, ``Width'' denotes the span, i.e., 
the shortest clear roof dimension across the excavated opening.
``Height'' denotes the vertical dimension. ``Length'' denotes the longitudinal length of elongated caverns (mainly tunnel-type caverns); for vertical silo-type caverns, ``Length'' is not applicable (see ($\dagger$)).
}
\label{tab:cavern_comp}
\centering
\begin{scriptsize}
\hspace*{-1.0cm}
\begingroup
\renewcommand{\arraystretch}{1.5}
\begin{tabular}{|c|c|c|c|c|c|c|c|c|c|c|}
\hline
\hline
Site & Country & Purpose & Shape & \multicolumn{3}{c|}{Dimensions [m]} & Volume & Overburden & Start of & Ref. \\ 
\cline{5-7}
     &      &      &      & Width & Height & Length & [m$^3$] & [m] & Operation & \\
\hline
\makecell{Hyper- \\ Kamiokande} & \makecell{Japan \\ (Gifu)} & \makecell{Particle \& \\ Astroparticle \\ Physics} & \makecell{Dome$+$ \\ Cylinder} & 69.0 & 93.7 & 
-- ($\dagger$)
  & 321,000 &  600 & 2025/7(*) & -- \\ 
\hline
\makecell{Super- \\ Kamiokande} & \makecell{Japan \\ (Gifu)} & \makecell{Particle \& \\ Astroparticle \\ Physics} & \makecell{Dome$+$ \\ Cylinder} & 40.0 & 57.6 & 
-- ($\dagger$)
  & 69,000  & 1,000 & 1996/4 & \cite{SK_excavation} \\ 
\hline
\makecell{Kazunogawa \\ Hydropower \\ Plant} & \makecell{Japan \\ (Yamanashi)} & Powerplant & \makecell{Egg- \\ shaped}   & 34.0 & 54.0 & 210.0 & 250,000 &  500 & 1999/12 & \cite{Kazunogawa} \\
\hline
\makecell{Baihetan \\ Hydropower \\ Plant} & China & Powerplant & \makecell{Warhead- \\ shaped}  & 34.0 & 88.7 & 438.0 & $1.3 \times 10^6$ (\#) & \makecell{260$\sim$ \\ 330} & 2022/12 & \cite{Baihetan} \\
\hline
\makecell{Cirata \\ Hydropower \\ Plant} & Indonesia & Powerplant & \makecell{Egg- \\ shaped} & 35.0 & 49.5 & 253.0 & 320,000 & 100 & 1988/9 & \cite{Cirata} \\
\hline
\makecell{Gj{\o}vik \\ Olympic \\ Cavern Hall} & Norway & \makecell{Sport \\ Stadium} & \makecell{Semi- \\ cylindrical} & 61.0 & 25.0 & 91.0 & 140,000 (\$) & \makecell{25$\sim$ \\ 50} & 1993/4 & \cite{Gjovik} \\ 
\hline
\makecell{Namikata \\ LPG \\ Storage} & \makecell{Japan \\ (Ehime)} & \makecell{LPG \\ Storage} & \makecell{Egg- \\ shaped} & 26.0 & 30.0 & \makecell{430.0$\sim$ \\ 485.0} & \makecell{281,000$\sim$ \\ 317,000} & 150 & 2013/3 & \cite{Namikata} \\
\hline
JUNO & China & \makecell{Particle \& \\ Astroparticle \\ Physics} & \makecell{Arch-shaped \\ Roof $+$ \\ Cylinder} & 45.6 & 71.9 & 45.6 & $\sim$120,000 & $\sim$650 & 2025/8 & \cite{JUNO} \\
\hline
DUNE & USA & \makecell{Particle \& \\ Astroparticle \\ Physics} & \makecell{Rectangular \\ arch-shaped} & \makecell{ 19.8 \\ 19.8 \\ 19.3} & \makecell{ 28.0 \\ 28.0 \\ 10.95} & \makecell{ 144.5 \\ 144.5  \\ 190} & \makecell{$\sim$78,000 \\ $\sim$78,000 \\ $\sim$36,000} & $\sim$1,500 & 2024/2(*) & \cite{DUNE} \\
\hline
CJPL & China & \makecell{Academic \\ Research} & \makecell{Rectangular \\ arch-shaped} & \makecell{ 14 \\ 14 \\ 14 \\ 14} & \makecell{ 14 \\ 14 \\ 14 \\ 14} & \makecell{ 130 \\ 130 \\ 130 \\ 130} & \makecell{ $\sim$25,500 (\#) \\ $\sim$25,500 (\#) \\ $\sim$25,500 (\#) \\ $\sim$25,500 (\#)} & 2,400 & 2023/12 & \cite{CJPL} \\ [5pt]
\hline
\hline
\multicolumn{11}{l}{\small (*) date of excavation completion, (\#) estimated by cuboid approximation, (\$) total volume of excavation,
} \\
\multicolumn{11}{l}{\small
($\dagger$) for vertical silo-type caverns, length is not applicable.
}
\end{tabular}
\renewcommand{\arraystretch}{1.0}
\endgroup
\end{scriptsize}
\end{table}

Large-scale underground rock caverns in Japan and abroad are typically constructed for underground power stations or storage facilities. Representative examples are compared in Table~\ref{tab:cavern_comp}~\cite{SK_excavation,Kazunogawa,Baihetan,Cirata,Gjovik,Namikata,JUNO,DUNE,CJPL}, complementing those in Fig.~\ref{fig:cavern_comp} (right). 
Mine cavities are excluded because their primary purpose is extraction and their stability requirements differ fundamentally from those of underground structures. Spaces formed by open-cut excavation were also excluded from consideration. Military facilities, for which information is not publicly available, are not considered. 

Many large rock caverns adopt tunnel-type layouts to maximize usable volume, with cross-sections whose spans are relatively small compared with their length. In such configurations, stability against in-situ stresses is governed primarily by the 
span---the shortest clear roof dimension across the excavated opening, (i.e., the roof clear span)---rather 
than by the overall length. 
Fig. ~\ref{fig:cavern_comp} (right) compares cross-sections by 
span-versus-height.
Baihetan in China, often cited as the ``largest underground power plant cavern,'' has a large total volume~\cite{Baihetan}, while the Gj{\o}vik Olympic Cavern Hall achieves a span of 61~m~\cite{Gjovik}. 
In contrast, HK and Super-Kamiokande (SK) employ a vertically oriented design with a dome-capped cylindrical shape, combined with substantial overburden to suppress cosmic-ray backgrounds. This configuration enables cost-effective construction, from cavern excavation through tank and detector installation. 
Among the caverns 
surveyed in this study,
HK exhibits the largest span and height. 

The HK project~\cite{HK-Koshiba,HK-Shiozawa,HK-Nakamura,HK-Itow,HK-Design} is an international collaboration led by the University of Tokyo and the High Energy Accelerator Research Organization (KEK). 
HK is intended to serve as a global flagship facility for particle and astroparticle physics for more than two decades. 

\begin{figure}[htbp]
\begin{center}
\includegraphics[width=0.8\linewidth]{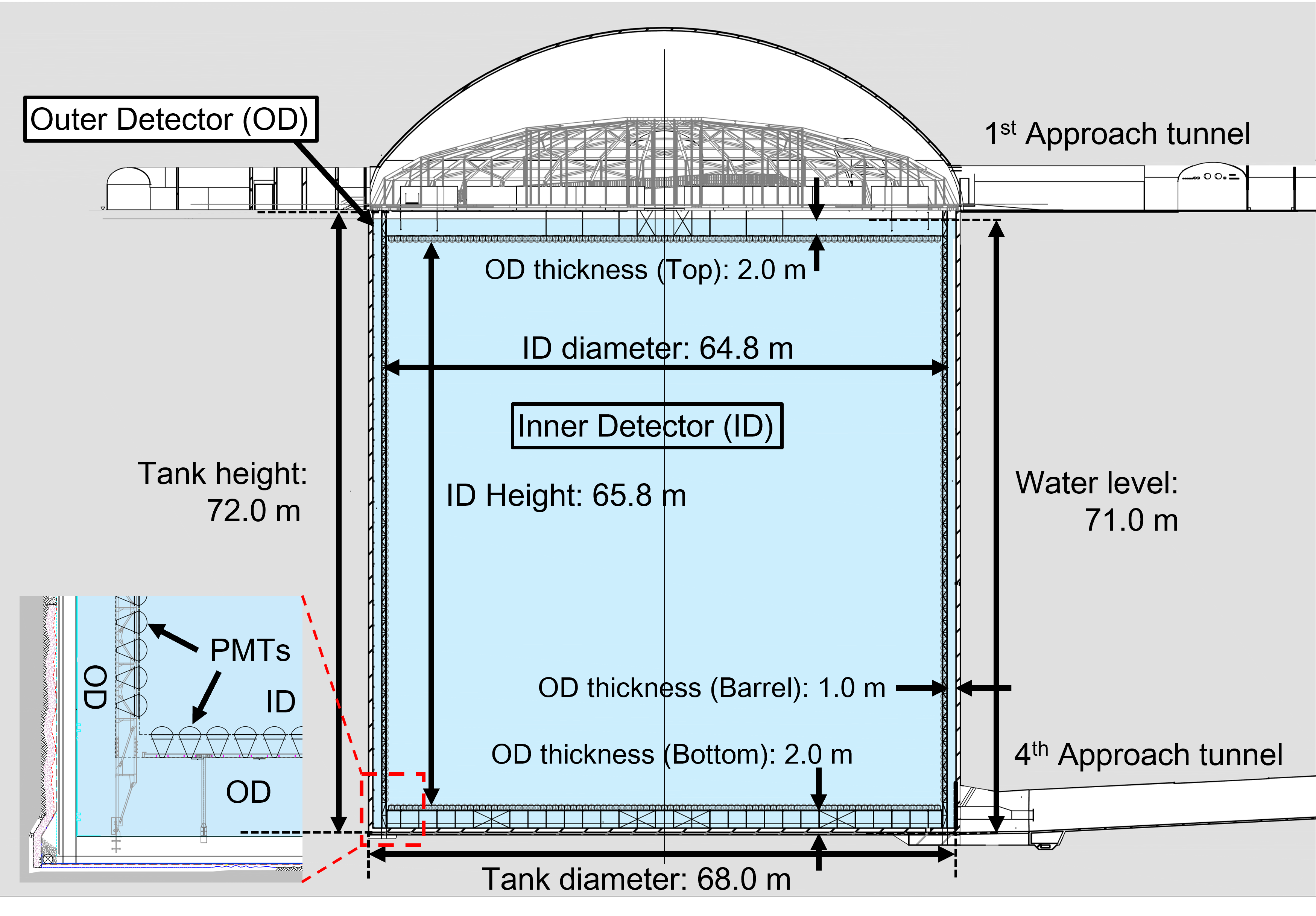}
\caption{Cross-section of the Hyper-Kamiokande stainless-steel water tank installed inside the 69~m-diameter cavern.
The tank is 68~m in diameter and 72~m in height and will be filled with ultra-pure water to 71~m (total 258,000~m$^3$; fiducial 188,000~m$^3$). The detector is partitioned into an Inner Detector (ID) and an Outer Detector (OD); about 20,000 pressure-tolerant 50-cm PMTs (expandable to $\sim$40,000) are mounted on the ID wall. 
This near-unity aspect ratio is a cost- and performance-driven detector choice that imposes stringent constraints on cavern excavation and stability. 
}
\label{fig:hk}
\end{center}
\end{figure}

HK will feature a water Cherenkov detector with a fiducial mass 8.4 times that of SK. 
The cavern will accommodate a cylindrical stainless-steel tank, 68~m in diameter and 72~m in height, which will be filled with ultra-pure water to a depth of 71~m (Fig.~\ref{fig:hk}). 
The total volume is 258,000~m$^3$, with a fiducial volume of 188,000~m$^3$ (corresponding to a fiducial mass of 188,000 metric tons). 
The tank is partitioned into inner and outer volumes by a stainless-steel structure, and approximately 20,000 50-cm photomultiplier tubes (PMTs) will be mounted on the inner wall facing inward (up to $\sim$40,000 PMTs can be installed) to detect Cherenkov photons in water. 
Taking advantage of the enhanced pressure tolerance of the PMTs~\cite{HKPMT}, which enabled an increased water depth, and given that photosensor coverage on the detector wall is a primary cost driver, the HK tank adopts a near-unity diameter--height aspect ratio to maximize the volume-to-surface-area ratio and achieve cost-effective performance. 

In addition to natural neutrinos from supernovae, the Sun, and the atmosphere, accelerator-produced neutrinos constitute a major pillar of HK. 
The geometrical configuration of the accelerator-based neutrino oscillation experiment is defined such that the Hyper-Kamiokande far detector site, located approximately 8~km south of SK, is positioned at an off-axis angle of 2.5$^\circ$ and at a baseline of 295~km with respect to the J-PARC neutrino beam~\cite{HK-Itow}. 
With the J-PARC beam power increased to 1.3~MW and the far detector enlarged, HK will record accelerator neutrinos at approximately 20 times the rate of the current Tokai-to-Kamioka (T2K) long-baseline neutrino experiment~\cite{T2K} (comparison based on the T2K beam power as of 2020), allowing precision measurements of the difference in oscillation probabilities between neutrino and antineutrino beams with overwhelming statistical significance~\cite{HK_LBL}. 
These measurements are expected to lead to the discovery of CP violation in neutrinos, which may provide critical insight into the origin of matter--antimatter asymmetry in the universe. 

Another primary objective is the most sensitive search for nucleon decay~\cite{HK-Design}, a generic prediction of grand unified theories (GUTs). Discovery would be direct evidence of GUTs, challenging the Standard Model and potentially triggering a paradigm shift in particle physics. The third pillar is precision observation of supernova neutrinos~\cite{HK_SNe}, aiming to elucidate supernova mechanisms and the history of star formation through measurements of supernova bursts and the relic supernova neutrino flux. 
For these physics goals, detector size---more precisely, the fiducial mass---primarily determines the sensitivity; enlarging the cavern was therefore a critical consideration in experimental design. 

As noted above, achieving precise observation of neutrinos---which only rarely interact with matter---and conducting the most sensitive search for yet-undetected nucleon decay require an underground detector with a very large fiducial mass, and hence a huge cavern. 
JUNO, a 20~kton liquid-scintillator neutrino detector located under approximately 650~m of rock overburden in Jiangmen, China, is inside a cavern larger than SK (see Table~\ref{tab:cavern_comp})~\cite{JUNO}. 
DUNE, 
a leading experiment competing with HK, 
is also excavating three enormous caverns~\cite{DUNE}, each comparable in scale to past underground power-station caverns. Likewise, for next-generation facilities such as direct dark-matter search experiments and double-beta decay experiments, underground spaces with extremely low background are indispensable, as in SK and HK. Consequently, excavation of large caverns has become an essential and non-negligible component of experimental planning. Table~\ref{tab:cavern_comp} also includes CJPL, a Chinese laboratory excavated at a depth of 2,400~m~\cite{CJPL}.

This paper documents the design and construction of the HK cavern excavation and is intended as a reference for future underground particle-physics experiments.
The structure of the paper is as follows: Sec.~\ref{sec:excavation} provides an overview of the underground works; Sec.~\ref{sec:geo} describes geological investigations; Sec.~\ref{sec:design} presents the final cavern design; Sec.~\ref{sec:info} discusses an information-based design and construction approach; and Sec.~\ref{sec:conclusion} concludes. Throughout this paper, Secs.~\ref{sec:excavation}--\ref{sec:design} and Sec.~\ref{sec:info} are tightly coupled. We present Sec.~\ref{sec:excavation} first to establish the underground facility layout and excavation sequence; the excavation plan was initially developed based mainly on the baseline geological understanding summarized in Ref.~\cite{HK-Design} and was subsequently refined using site-specific investigations and excavation-stage observations described in Sec.~\ref{sec:geo}. Secs.~\ref{sec:excavation} and \ref{sec:geo} together provide the primary inputs to the stability analysis and design framework summarized in Sec.~\ref{sec:design}, and the update process and its rationale (traceability from assumptions to observations to design revisions) are consolidated in Sec.~\ref{sec:info}. To keep the narrative focused on the overall picture, we present the final (as-built) configuration and parameter settings in Secs.~\ref{sec:excavation}--\ref{sec:design}. 

\section{Overview of Underground Works} \label{sec:excavation}
\begin{figure}[bth!]
\includegraphics[width=1.0\linewidth]{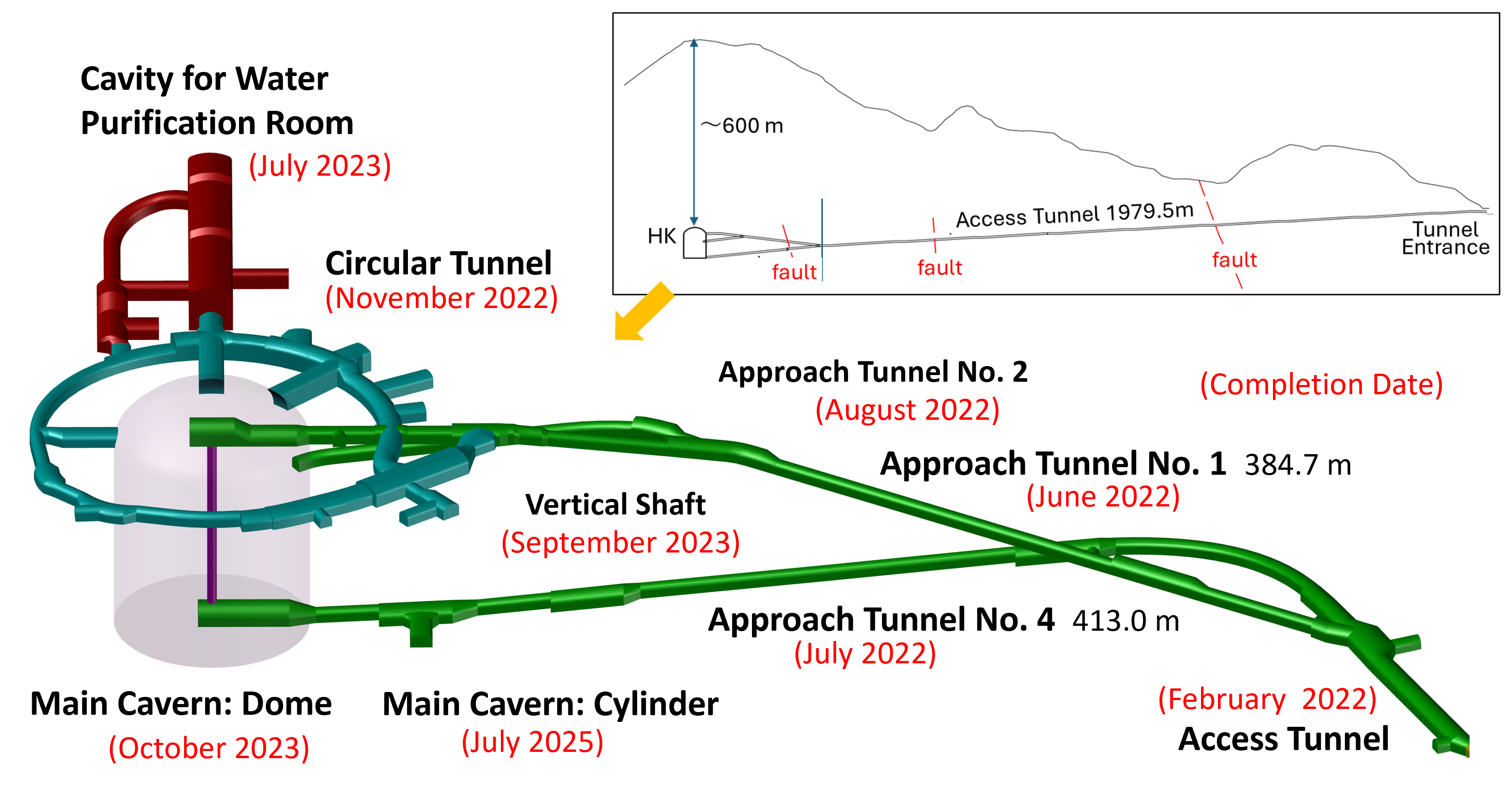}
\caption{Overview of the Hyper-Kamiokande underground facilities.
The total underground excavation for the project, including access tunnels and auxiliary caverns, is 
479,000~m$^3$. 
}
\label{fig:cavern_overview}
\end{figure}

To excavate the main cavern deep underground and install all equipment required for the experiment, the project included access tunnels and auxiliary caverns for pure-water systems and other subsystems in addition to the main cavern. Fig.~\ref{fig:cavern_overview} summarizes the HK underground facilities.
Including access tunnels and auxiliary caverns, the total underground excavation amounts to 
479,000~m$^3$. 

The HK project officially started in February 2020~\cite{HK_official_start}. Preparatory works were carried out during fiscal year 2020, and access-tunnel excavation began in May 2021. Owing to the hard rock, drill-and-blast methods were used. Each cycle comprised drilling, charging, blasting, mucking, scaling, shotcreting, and installation of rock bolts. Long-hole blasting achieved 3--4~m of advance per round, enabling rapid progress.

Subsequently, approach tunnels connecting to the top and bottom of the cylindrical section of the main cavern, a circular tunnel serving both geological investigations and auxiliary facility spaces, and caverns for pure-water systems responsible for producing and circulating ultra-pure water for the detector were excavated in sequence. Parallel operations were implemented wherever feasible to shorten the overall schedule. Completion dates for each tunnel and cavern are shown in Fig.~\ref{fig:cavern_overview}.

The main cavern consists of a dome section (69~m diameter, 21~m rise) designed to withstand $\sim$600~m of overburden, and a cylindrical section (69~m diameter, 73~m height) forming the detector tank. 
Excavation of the dome section of the main cavern began in November 2022, and the cylindrical section was completed in July 2025. 

\begin{figure}[!hbt]
\includegraphics[width=1.0\linewidth]{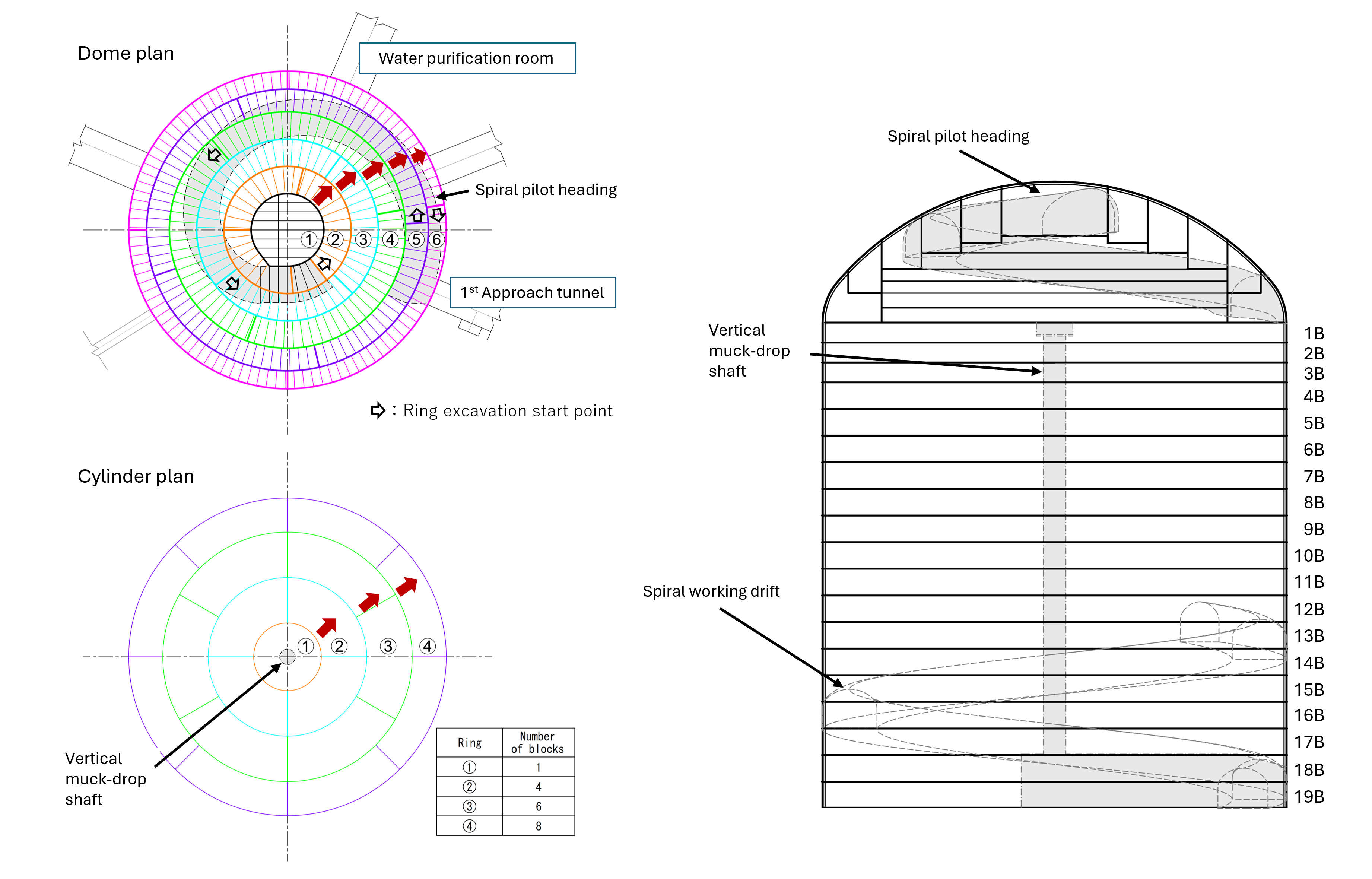} 
\caption{
Excavation sequence of the main cavern. 
The gray filled area in the dome indicates the spiral pilot heading (outlined with dashed lines), and the gray filled area in the cylinder indicates the vertical muck-drop shaft (outlined with dash-dotted lines). In the cylinder, the spiral working drift is shown only in the right panel as an open area outlined with dashed lines. 
Thick solid lines indicate ring and core boundaries in the dome, and bench boundaries in the cylinder; these boundaries coincide with PS-anchor installation boundaries.
(Right) Longitudinal cross-section encompassing both the dome and the cylinder. 
Thick solid lines indicate ring and core boundaries in the dome, and bench boundaries in the cylinder; these boundaries coincide with PS-anchor installation boundaries.
(Upper-left) Dome plan. Dome excavation proceeded circumferentially by rings, each starting from the boundary with the previously excavated spiral pilot heading (gray filled area). Thin solid lines indicate individual blast rounds. 
(Lower-left) Cylinder plan. Each bench was divided radially into four segments, and each segment was subdivided into nearly equal areas (blocks) for bench blasting (thin solid lines). 
}
\label{fig:division}
\end{figure}

Fig.~\ref{fig:division} shows the excavation sequence (advance-support division) for the dome and cylinder. The right panel presents a longitudinal cross-section encompassing both; the upper-left panel shows the dome plan; and the lower-left panel shows the cylinder plan. 
The dome excavation started with 
a spiral pilot heading to reach the crown, followed by ring-wise enlargement. 
The cylinder was benched downward in 3-4~m lifts (19 benches total, 1B--19B), with muck dropped into a vertical muck-drop shaft and hauled by dump trucks via Approach Tunnel No. 4 (see Fig.~\ref{fig:cavern_overview}). 
To mitigate potential issues and to confirm geological conditions inside the cylindrical section, a spiral working drift from Approach Tunnel No. 4 into the cylindrical section was also excavated first. 

In the dome, excavation proceeded circumferentially by rings (6 rings total, 1R--6R); after advancing $\sim$1/3--1/4 of a ring, prestressed (PS) anchors and instruments were installed
(see Fig.~\ref{fig:app_excavation_procedure} for a schematic overview).
Each ring started from the boundary with the previously excavated spiral pilot heading. 
Thin
solid lines indicate individual blast rounds; thick solid lines indicate PS-anchor installation boundaries; the gray filled area denotes the spiral pilot heading (outlined with dashed lines). 
Because instruments begin recording only after installation, their data capture post-installation increments rather than deformation that occurred during prior excavation. Accordingly, the sequential stability analyses (Sec.~\ref{sec:design}) were synchronized with the installation schedule and evaluated at matching step boundaries. 

In the cylinder, each bench was divided radially into four segments, and each segment was subdivided into nearly equal areas (blocks) for bench blasting as shown by the thin solid lines in the lower-left panel of Fig.~\ref{fig:division}. 
By scheduling blasting at shift-change times, two-shift operations eliminated the need for separate blast-withdrawal and subsequent waiting periods, thereby improving excavation efficiency. Support was required only at the outer periphery of the cylindrical section and consisted of primary shotcrete, rock bolts, secondary shotcrete, and PS anchors. To improve efficiency, these support operations were performed in parallel with core blasting up to the third ring. 

In specific cases, the sequence was modified. 
When the spiral working drift (see dashed lines in the right panel of Fig.~\ref{fig:division}) intersected the bench blasting blocks, long-hole blasting with extended drill lengths was combined with trimming blasts to level bench floors, changing the sequence from Bench 
11 (11B)
onward 
(see Fig.~\ref{fig:app_barrel_division} for the as-built bench partitioning from 11B to 19B).
In addition, because the rock mass above Approach Tunnel No.~4 proved poorer than anticipated, the breakthrough---originally scheduled at 17B with a 4~m rock web---was brought forward and executed at 16B by long-hole blasting with an 8~m rock web; consequently, from 16B onward the excavation sequence, especially the bench partitioning in the affected sections, was significantly revised.

Throughout the project, excavated muck was transported out of the underground facilities by dump trucks, appropriately sorted and managed, and was sent for reuse as embankment material or as aggregate for concrete. Groundwater inflow during excavation was limited and was properly collected and treated on site before discharge.

PS anchors, together with shotcrete, serve as long-term support elements ensuring cavern stability. Rock bolts were installed as face-support elements to maintain safety during advance but were not considered in long-term stability calculations. 
A PS anchor consists of a seven-wire prestressing steel strand, an anchorage (fixed) zone, a free length covered by polyethylene sheathing, and a head assembly with bearing plate 
(see Fig.~\ref{fig:app_ps_anchor_structure} for the structural details of the PS anchor). 
The annulus between the steel and the sheath is filled with grease to allow sliding within the rock mass. Anchors are installed in drilled holes and grouted with mortar; the fixed zone develops in competent rock at depth, while the sheathed free length accommodates relative movement between the head and the fixed zone. 
Initial tension applied at the head actively restores part of the confining force lost due to excavation. Drawing on experience from underground power-station projects, PS anchors constitute the critical support element for long-term stability 
(see Fig.~\ref{fig:app_bird_ps_1} and \ref{fig:app_bird_ps_2} for bird's-eye views of the cavern including all the PS anchors installed).

\section{Geological Investigations} \label{sec:geo}
The HK site lies within the Hida Belt, a composite terrane of metamorphic (Hida gneiss) and granitic rocks, broadly divided into eastern, central, and western domains with distinct lithologies and structures. HK is in the central domain; the site is primarily metamorphic rock with coarse-grained, massive, igneous-like textures (e.g., dioritic gneiss)~\cite{kamioka_mine}.
Overall, the site conditions appear highly favorable by Japanese standards for large underground cavern excavation, with the basis set out below. 

Site selection employed extensive investigations: exploratory drilling, tunnel-wall mapping, initial rock-stress measurements, and elastic-wave surveys. These results were consolidated into a geological model, representing observed lithology, weak-layer continuity, and rock-mass classification (ranks B, CH, CM, and CL based on fracture spacing and alteration)~\cite{rockclass}. From most competent to less competent, ranks are B, CH, and CM; CL denotes rock with predominant fracturing and partial clayey zones
(see Table~\ref{tab:app_rock_mass_classification} for more detail).
Investigations confirmed that the HK cavern is hosted mainly in high-quality B-rank rock, with no throughgoing faults that would critically affect stability~\cite{HK-Design} 
(for the regional geological map, see Fig.~\ref{fig:app_wide_geo_map}). 

\begin{figure}[!hbtp]
\begin{center}
\includegraphics[width=1.0\linewidth]{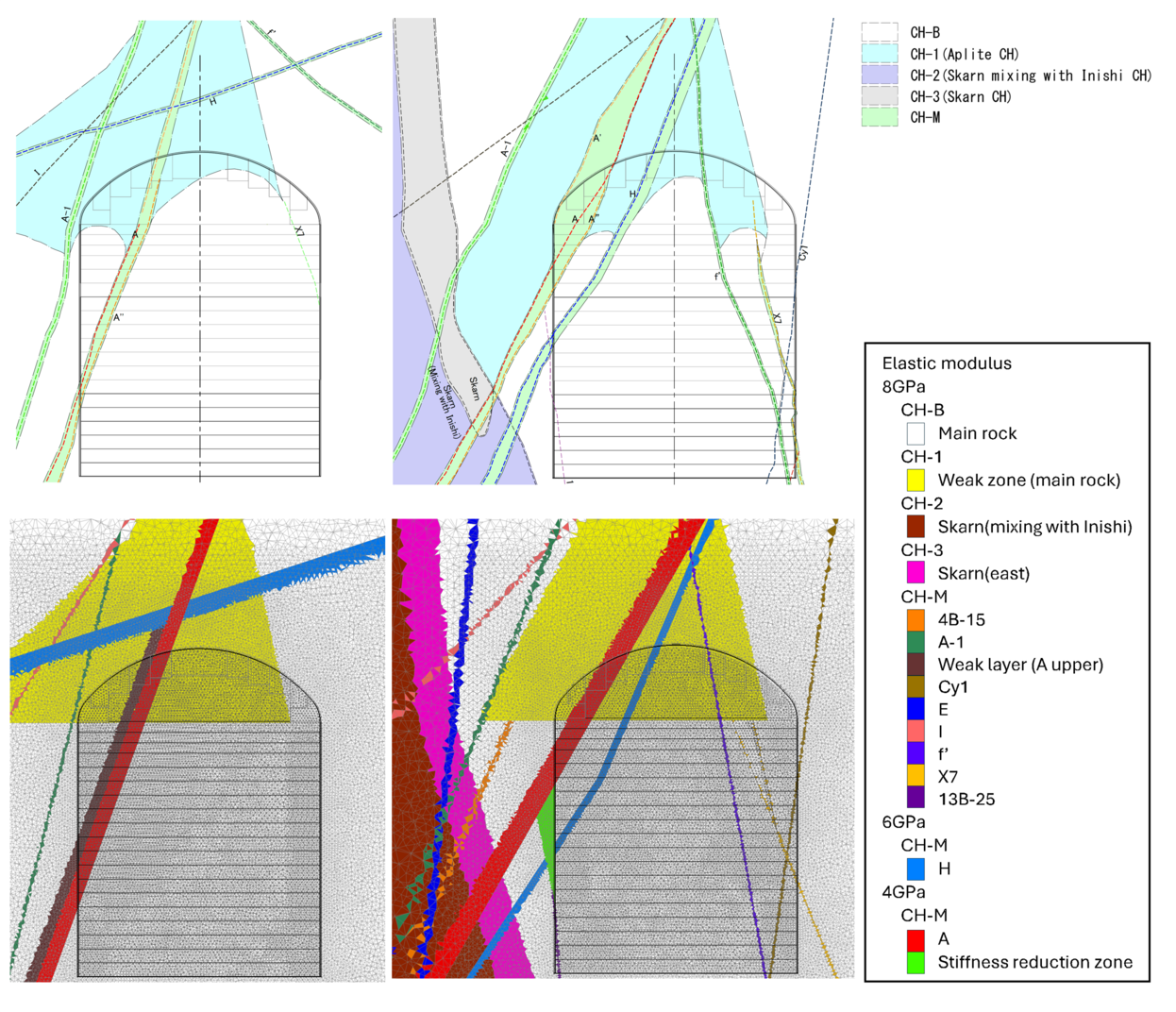}
\caption{\small
Geological and analysis models.
Upper panels: final geological model (left: north--south section; right: east--west section) developed from pre-construction investigations and subsequently updated through the information-based (observational) design and construction approach.
Lower panels: stability-analysis model (left: north--south; right: east--west) incorporating the analytical geological model (simplified geological classifications, planar weak-layer representation, and parameterized rock-mass properties) together with excavation geometry and boundary conditions.
Legends indicating rock-mass classes and weak layers are shown within the figure.
Rock-mass class labels used in the geological model denote grouped/mixed geological units based on observations: CH-B (CH--B mixed), CH-1 (aplite--CH mixed), CH-2 (skarn--Inishi mixed), CH-3 (skarn CH), and CH-M (CM and CM-like CH).
}
\label{fig:cavern_geo}
\end{center}
\end{figure}

In fiscal year 2020, large-scale investigations finalized the cavern location and design conditions. 
Two investigation (sampling) adits (total length 96~m) were excavated from existing Kamioka Mine tunnels---one heading toward the planned dome center and another for skarn~\cite{skern} characterization. 
In addition, boreholes totaling 725~m were drilled both upward from a point near the planned dome center at the base of this adit and downward from higher levels, enabling detailed surveys of the geology above the dome crown, the continuity of weak layers, rock-mass strength and deformability, and initial rock stress. 
(The relative positions of boreholes, adits, and the cavern are also shown in Fig.~\ref{fig:app_wide_geo_map}).
Key items are summarized below (see Fig.~\ref{fig:cavern_geo}, upper panels, which present the final geological model consolidated after construction began and updated through the information-based (observational) design and construction approach described in Sec.~\ref{sec:info}). 

\paragraph{\bf Groundwater Conditions}
Despite the $\sim$600~m overburden, the groundwater table lay below the planned cavern floor, and water inflow during excavation was negligible, minimizing dewatering requirements and facilitating smooth excavation progress. 
The HK site occupies a part of the Kamioka Mine long unused due to sparse ore veins; existing tunnels in the vicinity likely acted as preferential drainage pathways.

\paragraph{\bf Initial Rock Stress}
Initial stress is a critical design parameter and varies by region; in situ measurements are essential. At HK, stress was measured using the conical-ended borehole method~\cite{inistress}. Of seven measurement points, those suspected of stress relief due to fractures and those showing vertical stress far exceeding the overburden estimate ($\sim$15.8~MPa) were excluded, 
as such values are unlikely to be representative over the cavern scale. 
Only a limited number of cases was used because the goal was not to envelope all possible combinations of principal stresses, but to bracket the behaviors most relevant to stability and support design with a nominal case and a small number of credible risk cases. We therefore defined one nominal case and a small number of representative risk cases, separately for the dome and the cylinder, based on the measured stresses and their relevance to stability. 

The average of the remaining three points was adopted as the nominal case (Case A). The maximum principal stress was $\sigma_1$ = 15.1~MPa, consistent with the overburden estimate; however, the principal-stress ratio was high, $\sigma_1 / \sigma_3 = 3.69$, characteristic of the HK site. 
During dome excavation, where $\sigma_1$ acts perpendicular to the eastern dome wall, tensile failure along the $\sigma_3$ direction may occur. 
The measurement point with the highest stress ratio ($\sigma_1 / \sigma_3 = 5.4$) was designated Case C and used as a risk case for the dome. For reference, SK also exhibited a high stress ratio (4.7)~\cite{SK_excavation}.
In the cylindrical section, shear failure predominates. Higher horizontal stress amplifies secondary stress uniaxialization (i.e., stress redistribution becoming locally one-dimensional) and enlarges the loosened zone. 
Including points where vertical stress exceeded the overburden estimate, the average of five locations was considered (Case D). Because widespread stress significantly exceeding overburden is unlikely, Case D was adjusted by a factor of 1/1.2 to define Case D' for the cylinder risk case. Table~\ref{tab:init_stress} summarizes these values
(see Fig.~\ref{fig:app_inist_dir} for the principal stress orientations and magnitudes).
\begin{table}[!h]
\caption{Summary of initial rock-stress values.
Stress components are resolved in a site-fixed Cartesian coordinate system with $+$X east, $+$Y north, and $+$Z upward. 
$\sigma_{x}$, $\sigma_{y}$, $\sigma_{z}$ are normal stresses; $\tau_{xy}$, $\tau_{yz}$, $\tau_{zx}$ are shear stresses. $\sigma_1$, $\sigma_2$, $\sigma_3$ denote principal stresses. 
Compressive stresses are taken as positive. The values reflect an overburden of approximately 600~m, corresponding to a reference vertical stress of 15.8~MPa. 
Note that the label `Case B' is not used in this paper. 
}
\label{tab:init_stress}
\centering
\hspace*{-0.5cm}
\begin{tabular}{|c|ccc|ccc|ccc|c|}
\hline
Initial rock & $\sigma_{x}$ & $\sigma_{y}$ & $\sigma_{z}$ & $\tau_{xy}$ & $\tau_{yz}$ & $\tau_{zx}$ & $\sigma_1$ & $\sigma_2$ & $\sigma_3$ & $\sigma_1/\sigma_3$\\
stress & [MPa] &  [MPa] &  [MPa] &  [MPa] &  [MPa] &  [MPa] &  [MPa] &  [MPa] &  [MPa] & \\
\hline
{\bf Case A} & 9.19 & 7.87 & 15.06 & -0.68 & -0.49 & 6.27 & 19.10 & 7.85 & 5.17 & 3.69 \\
{\bf Case C} & 7.43 & 8.14 & 15.04 &  0.61 & 1.64 & 6.66 & 19.18 & 7.89 & 3.54 & 5.42 \\
{\bf Case D} & 18.45 & 12.46 & 19.47 & -1.42 & -0.53 & 6.12 & 25.24 & 13.29 & 11.84 & 2.13 \\ 
{\bf Case D'} & 15.38 & 10.38 & 16.23 & -1.18 & -0.44 & 5.10 & 21.03 & 11.08 & 9.87 & 2.13 \\ 
\hline
\end{tabular}
\end{table}

\paragraph{\bf Rock-Mass Properties}
Rock-mass properties used for design are summarized in Table~\ref{tab:geo_params}. Key values were verified in situ. For B-class rock predominant at HK, the elastic modulus was determined by plate-loading tests~\cite{geo_tests}, and the shear strength by rock-mass shear tests~\cite{geo_tests}. Shear strength was evaluated in terms of peak and residual parameters; cohesion and friction angle were derived using the Mohr-Coulomb criterion\cite{standard}. Parameters for CH and CM classes were estimated using the Geological Strength Index (GSI) system~\cite{GSI}, while CL-class parameters were taken from representative literature values~\cite{CL}. Because direct measurement of tensile strength of rock mass is difficult, tensile strength was inferred from the Hoek-Brown criterion~\cite{Hoek} within the GSI framework 
(see Appendix~\ref{sec:app_gsi}). 
Alternative estimates (e.g., based on empirical correlations between uniaxial compressive strength and tensile strength or cohesion-tensile correlations) were considered, but the conservative lowest value obtained from Hoek-Brown was adopted, consistent with prior practice~\cite{tensile_refs}. 
Intact rock strength was measured in the laboratory and used to define the properties of aplite~\cite{aplite} in the dome. 

\begin{table}[!h]
\caption{Rock-mass properties compiled in this study; the section-specific parameter settings adopted in the final analysis models are given in Tables~\ref{tab:dome_geo} and \ref{tab:barrel_geo}. 
Here, $C_{\rm peak}$ and $\Phi_{\rm peak}$ are the peak cohesion and peak friction angle, and $C_{\rm res.}$ and $\Phi_{\rm res.}$ are the corresponding residual values in the Mohr--Coulomb shear-strength model. 
}
\label{tab:geo_params}
\centering
\begin{tabular}{|c|c|c|c|c|c|c|c|c|c|}
\hline
\multicolumn{2}{|c|}{Rock $\cdot$} & Unit & Elastic  & Poisson's & \multicolumn{4}{c|}{Shear strength} & Tensile \\
\cline{6-9}
\multicolumn{2}{|c|}{Rock mass}    & weight & modulus & ratio & \multicolumn{2}{c|}{Peak} & \multicolumn{2}{c|}{Residual} & strength \\
\cline{3-10}
\multicolumn{2}{|c|}{\makecell{classification \\ }} & $\rho$ & E & $\gamma$ & $C_{\rm peak}$ & $\Phi_{\rm peak}$ & $C_{\rm res.}$ & $\Phi_{\rm res.}$ & $\sigma_{t}$ \\
\multicolumn{2}{|c|}{ } & [kN/m$^3$] & [GPa] & & [MPa] & [deg] & [MPa] & [deg] & [MPa] \\
\hline
 & B & 27 & 40 & 0.25 & 5.4 & 65 & 0.7 & 53 & 0.7 \\
\cline{2-10}
\makecell{Inishi \\ Gneiss} & \makecell{CH \\ CH-8G} & 27 & \makecell{25 \\ 8} & 0.25 & 3.1 & 56 & 0.5 & 48 & 0.4 \\
\cline{2-10}
 & \makecell{CM \\ CM-6G \\ CM-4G} & 27 & \makecell{8 \\ 6 \\ 4} & 0.25 & 1.2 & 48 & 0.4 & 40 & 0.1 \\
\cline{2-10}
& CL & 27 & 0.7 & 0.25 & 0.7 & 34 & 0.7 & 34 & 0.0 \\
\hline
Skarn & \makecell{B \\ CH} & 27 & \makecell{ 14 \\ 8} & 0.25 & \makecell{2.3 \\ 1.3} & \makecell{64 \\ 51} & \makecell{0.6 \\ 0.5} & \makecell{51 \\ 42} & \makecell{0.3 \\ 0.1}\\ 
\hline
Aplite & \makecell{CH \\ CH-8G} & 27 & \makecell{16 \\ 8} & 0.25 & 2.5 & 56 & 0.5 & 45 & 0.3 \\
\hline
\end{tabular}
\end{table}

\begin{figure}[!bth]
\begin{center}
\includegraphics[width=1.0\linewidth]{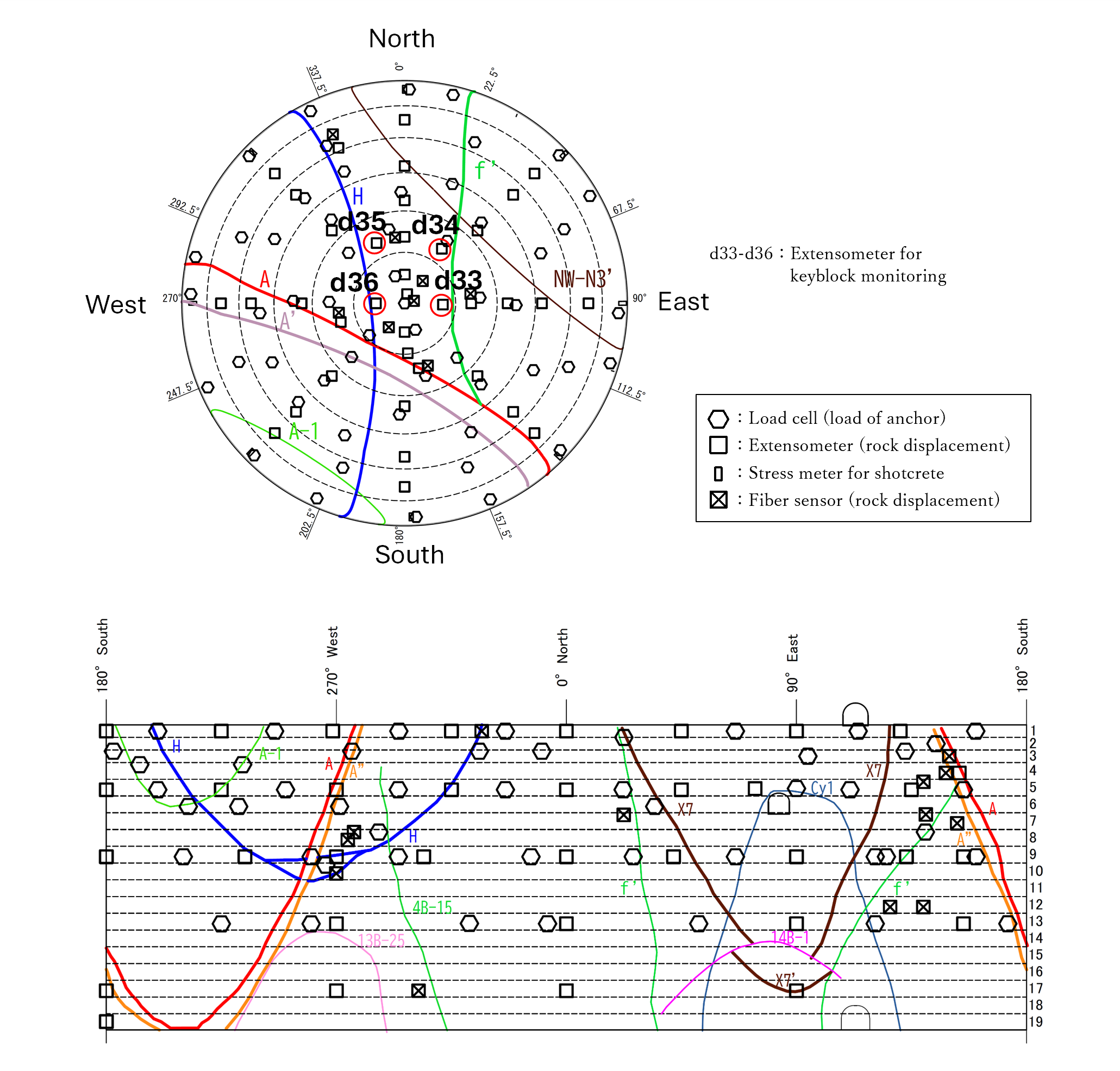}
\caption{\small
Distribution of weak layers and instrument layout. Upper: dome; lower: cylinder. 
Colored lines indicate the weak layers listed in Tables~\ref{tab:dome_geo} and \ref{tab:barrel_geo}. 
Thick lines indicate Weak Layers A, H, f', and X7, which are critical to cavern stability. 
Hollow hexagons indicate PS-anchor load cells; hollow squares, extensometers; hollow vertical rectangles, shotcrete stress gauges; hollow diamonds, 
fiber-optic displacement meters~\cite{fiber_displacement_sensor}. 
Concentric dashed rings in the dome represent excavation stages (1R--6R from the center); 
horizontal dashed lines in the cylinder represent bench boundaries (1B--19B from the top). 
}
\label{fig:cavern_wl}
\end{center}
\end{figure}

\paragraph{\bf Weak Layers}
Although no major faults are present at the HK site, smaller-scale discontinuities cannot be avoided. Highly continuous stratiform discontinuities associated with locally reduced rock-mass classification were identified as weak layers and treated as key monitoring targets. In hard rock, weak layers are a primary driver of deformation and are essential inputs for stability analysis. Initial identification, based on boreholes and borehole-wall observations, revealed highly continuous features such as Weak Layers A and H, characterized by clay intercalations and fracture concentration. 
Their location, strike, dip, and continuity (see Fig.~\ref{fig:cavern_wl}) were initially assessed during model development, including features with uncertain continuity, and later refined through tunnel-wall mapping of the circular tunnel. 
Continuity at the cavern scale cannot be uniquely determined from borehole inspection alone and was refined using tunnel-wall mapping and excavation-stage observations. 
As excavation progressed, geological maps were iteratively updated using drilling-energy data obtained during drilling of measurement boreholes and boreholes for PS anchors~\cite{MWD}, borehole television (BTV), and face/wall inspections. 
Drilling-energy data were used as a quantitative indicator to calibrate rock-mass classification and to check consistency of weak-layer orientation.
For simplicity of exposition, we treat two geologically major discontinuities that are critical to cavern stability as weak layers; hence, alongside Weak Layers A and H, 
we have two discontinuities denoted as Weak Layer f' and Weak Layer X7. 
All identified weak layers and major discontinuities exposed on the cavern walls are shown in Fig.~\ref{fig:cavern_wl}: the upper panel for the dome and the lower for the cylinder 
(see Fig.~\ref{fig:app_bird_inst_wl} for bird's-eye views of the cavern illustrating the relationship with the major Weak Layers).

\section{Cavern Design} \label{sec:design}
Excavation creates an empty space in rock mass that was previously in equilibrium under initial stress determined largely by overburden. 
Stresses formerly carried by the excavated volume redistribute into the surrounding rock, inducing deformation. 
Where stresses resulting from redistribution exceed rock strength, failure occurs, forming a loosened zone that may lead to instability. 
For a cavern with a large span under substantial overburden, such instability can be pronounced. 
Rock fails in shear and tension; unlike metals, rock is notably weak in tension. Initial stress is governed not only by overburden but also by geological history and topography~\cite{geo_hist}; variations in magnitude and principal components significantly affect cavern stability. 
In design, initial stress and geological information were first consolidated into a geological model. This was then abstracted into an analytical geological model for stability analysis, involving simplification of geological classifications, planar representation of weak layers, and parameterization of rock-mass classes. The complete stability-analysis model combined this analytical geological model with excavation geometry, boundary conditions, and constitutive laws, enabling sequential excavation simulations to estimate loosened zones. Based on these, the quantity and arrangement of support---shotcrete and PS anchors---are designed to stabilize potential failure regions. Consistency between model predictions and measurements during excavation is essential and is a core feature of the HK cavern design.

Japan's experience in analyzing large-cavern excavations dates back to underground power-station projects more than two decades ago; since then, domestic analytical activity in this area remained limited until HK. 
Those earlier designs typically relied on 2D analyses with tunnel-like geometries to reduce computational demand~\cite{anal_2D}. In contrast, the cylindrical geometry of HK combined with non-axisymmetric in situ stress required full 3D analysis. Accordingly, an elasto-plastic finite-difference analysis using FLAC3D (v7) was adopted.

The models comprised $\sim$9 million elements in a domain five times the cavern dimensions. Mesh refinement around the cavern used minimum element sizes of $\sim$0.7~m at the wall and $<$1.5~m within 25~m of the wall. Fixed boundary conditions were imposed on all six faces. Initial stress measured in situ was scaled element-wise by the ratio of local overburden to 
overburden 
at the measurement location (ground elevation EL 1,150.5 m; overburden $\sim$600 m). 
The Mohr-Coulomb criterion~\cite{standard} was used to identify plastic (loosened) zones. 
From this equal-stress reference state---namely, the prescribed in-situ stress field assigned element-wise and scaled by local overburden---we advanced the solution stepwise and evaluated failure against the current stress state in each element at each step, with any plastic yielding and associated stress redistribution taken into account. 
The constitutive model assumed perfect elasto-plasticity: after shear failure beyond peak strength, stresses were redistributed to avoid exceeding peak parameters; after tensile failure, the material was treated as no-tension, redistributing stresses to eliminate tensile components. In the dome, loosened zones with potential instability were stabilized by PS anchors tying into competent rock above, in combination with shotcrete (suspension of unstable blocks). In the cylinder, arbitrary slip surfaces may form; stability was ensured by resistance from PS anchors and shotcrete.

Rock mass generally exhibits strain-softening after failure~\cite{StrainSoftening}. In the present analysis, tensile failure was treated with strain-softening behavior as described above, whereas shear failure was assumed to retain peak strength even after failure, resulting in an elasto-plastic approach. Tensile failure dominated loosened zones in the dome; shear dominated in the cylinder. Consequently, elasto-plastic modeling without shear softening tends to underestimate loosened zones in the cylinder. This was mitigated by carefully modeling macroscopic discontinuities and calibrating against observed dome behavior, achieving practical predictive accuracy. 
The effect of PS-anchor pretension on suppressing loosened zones was also examined. For the pretension applied in this project ($\lesssim$ 50~kPa; for example, taking a nominal tributary area of $3 \times 4$m$^2$, 
600- and 300-kN pretensions correspond 
to equivalent wall pressures of 50 and 25~kPa, respectively), no significant suppression effect was observed. This result is reasonable because the pretension is less than 1~\% of the minimum principal stress ($\sigma_3$). 

\begin{figure}[!h]
\begin{center}
\includegraphics[width=1.0\linewidth]{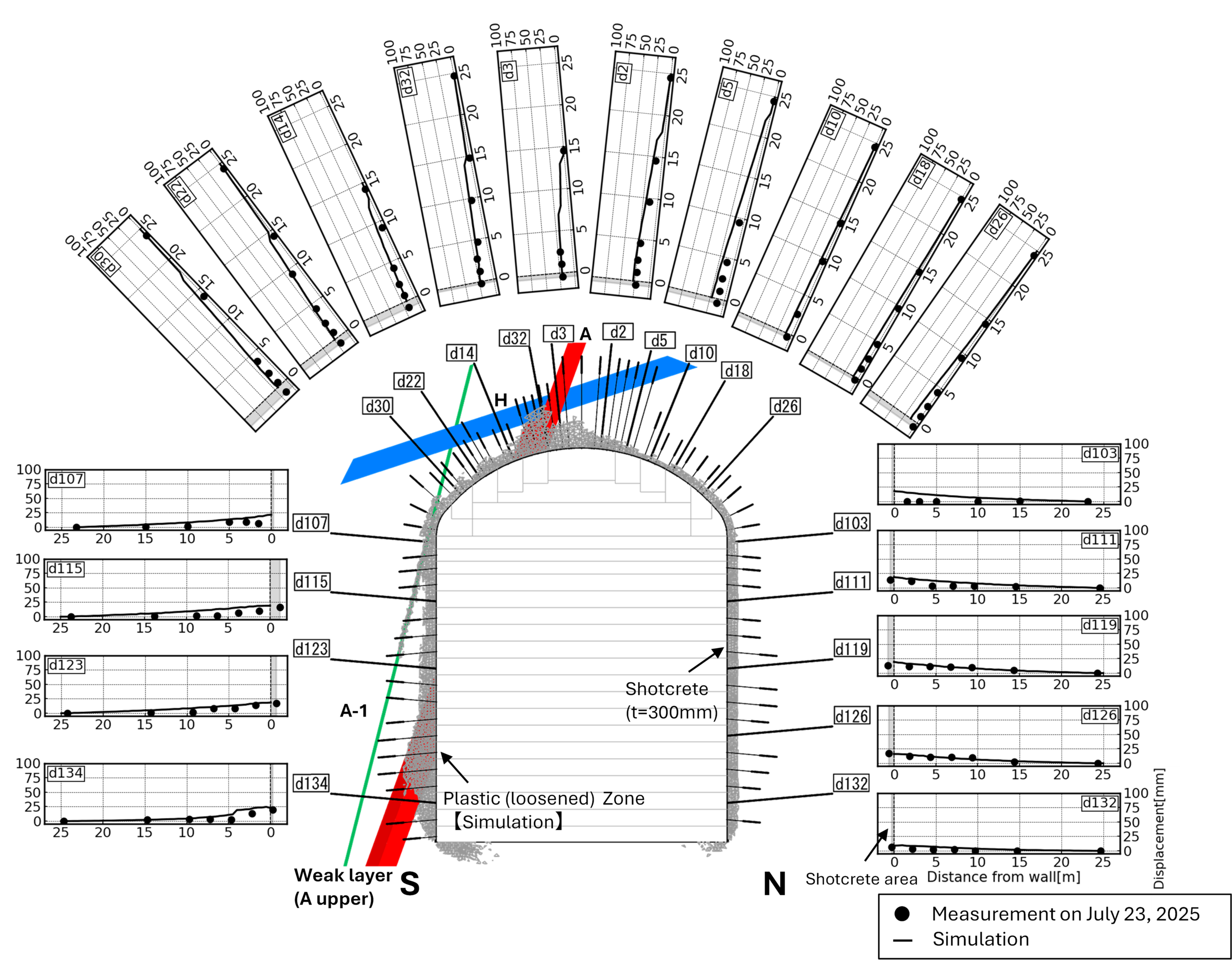}
\caption{
North--south section of the cavern showing plastic (loosened) zones (gray), mapped weak layers, support arrangement, and instrument layout.
PS anchors are indicated by thin lines (free length) and thick lines (fixed length), demonstrating that anchors are bonded in competent rock beyond the plastic zone.
Displacement profiles at completion compare model predictions (solid lines) with measurements (black circles), including instruments intersecting major weak layers, confirming consistency between the analysis and observed cavern-scale deformation. 
}
\label{fig:supportNS}
\end{center}
\end{figure}

\begin{figure}[!h]
\begin{center}
\includegraphics[width=1.0\linewidth]{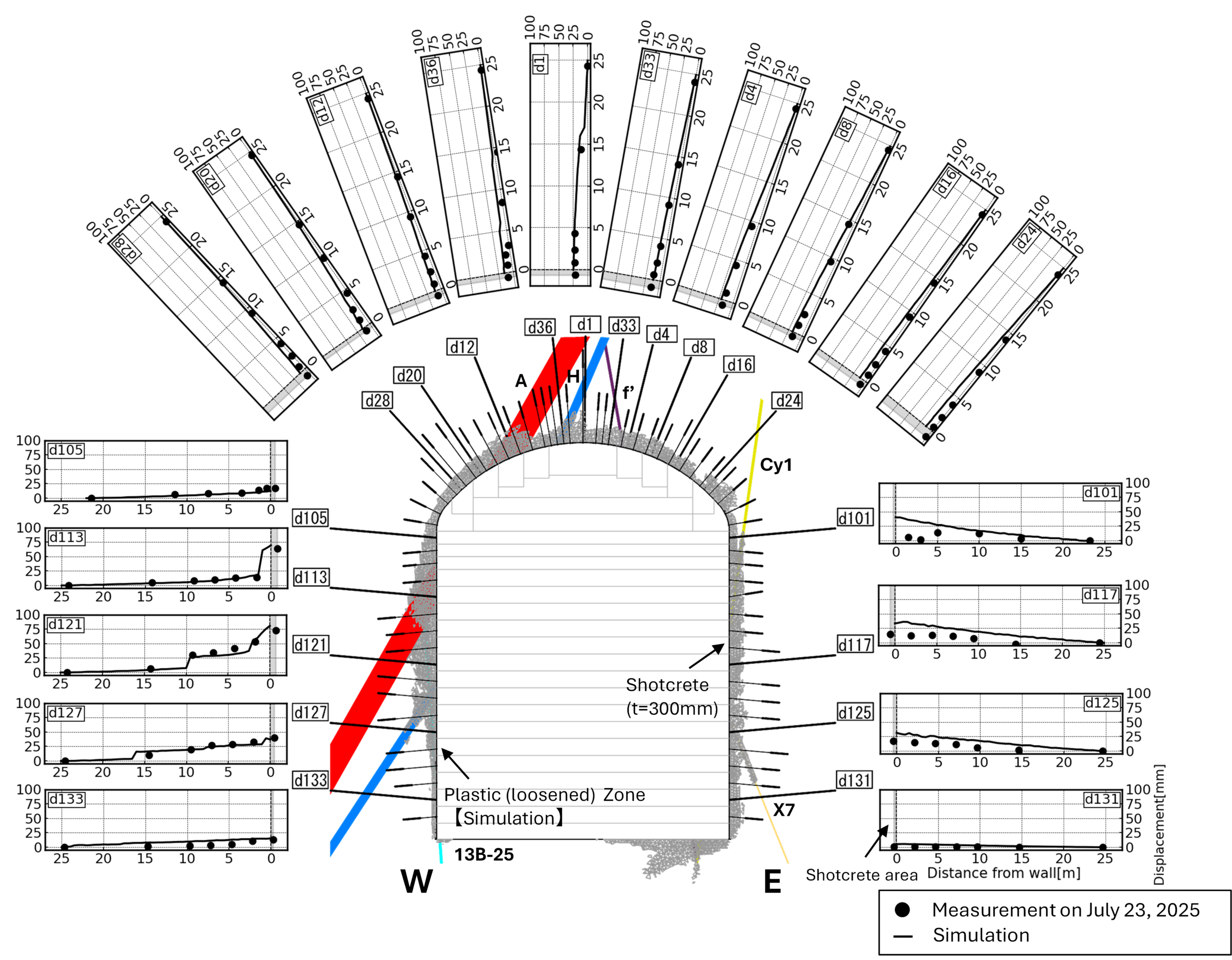}
\caption{
East--west section of the cavern. Plastic (loosened) zones (gray), weak layers, support arrangement, and instrument layout are shown, analogous to Fig.~\ref{fig:supportNS}. PS anchors are indicated by thin lines (free length) and thick lines (fixed length). Displacement profiles at completion compare predictions (solid lines) with measurements (black circles), including instruments intersecting major weak layers, demonstrating consistency between the analysis and observed deformation. For stability calculations, PS anchors whose fixed (bonded) length lies within the predicted plastic (loosened) zone are not counted.
}
\label{fig:supportEW}
\end{center}
\end{figure}

Using 2020 investigation results (rock-mass properties, in situ stress, and highly continuous weak layers), sequential excavation analyses aligned with the construction plan were performed to estimate loosened zones and design supports. Final results are presented in Figs.~\ref{fig:supportNS} (representative north--south
section) and \ref{fig:supportEW} (east--west
section), comparing plastic (loosened) zones, support arrangements, instrument layouts, and measured displacements with model predictions. 
Although models and support designs were updated during excavation through the information-based approach (Sec.~\ref{sec:info}), this section summarizes the final model; dome and cylinder are treated separately.
Weak layers were idealized as planar surfaces; interface elements and zones of reduced classification were defined per geology. Strike/dip/position were revised as needed during excavation. Material-property summaries for dome and cylinder models are provided in Tables~\ref{tab:dome_geo} and \ref{tab:barrel_geo}. Fig.~\ref{fig:cavern_geo} (lower panels) shows the analysis models with these approximations. Coordinates were set with origin 73~m below the dome floor (EL 550.5~m); $+$X east, $+$Y north, $+$Z upward.

\begin{table}[!h]
\caption{Final parameter settings for weak layers and geological units in the dome section.
These parameters define the section-specific input set used in the 3D sequential elasto-plastic analyses. 
Strike and dip describe the orientation of a planar feature: strike is the azimuth of a horizontal line on the plane, and dip is the angle of maximum downward inclination measured perpendicular to strike. We report strike in quadrant notation (e.g., N60W) and dip as dip angle with dip direction (e.g., 73S); for example, N60W73S denotes a plane with strike azimuth 300$^\circ$ and dip 73$^\circ$ toward south. 
Here, $K_n$ and $K_s$ denote the normal and tangential spring constants of the interface (IF) elements, respectively. 
}
\label{tab:dome_geo}
\centering
\begin{small}
\begin{tabular}{|c|c|c|c|c|c|c|c|c|}
\hline
\multicolumn{3}{|c|}{Elements of Model} & Width & Rock mass & Strike & \multicolumn{3}{c|}{Interface Element} \\
\cline{7-9}
\multicolumn{3}{|c|}{} & [m] & classification & and dip & Position & $K_{n}$ & $K_{s}$ \\
\multicolumn{3}{|c|}{} & &  & &  & {\scriptsize [GPa/m]} & {\scriptsize [GPa/m]} \\
\hline
\multicolumn{3}{|c|}{Main Rock} & -- & \makecell{Aplite \\ CH-8G} & -- & -- & -- & -- \\
\hline
\multicolumn{2}{|c|}{ } & A & 3.5 & \makecell{Inishi \\ CM-4G} & N60W73S & -- & 70 & 0.7 \\
\cline{3-9}
\multicolumn{2}{|c|}{\makecell{\hspace{1cm} \\ Weak Layer}} & A upper & 2.5 & Inishi CM & N60W73S & \makecell{ cylinder \\ east}  & -- & -- \\
\cline{3-9}
\multicolumn{2}{|c|}{ } & H & 2 & \makecell{Inishi \\ CM-6G} & N8W67W & -- & 70 & 0.7 \\
\cline{3-9}
\multicolumn{2}{|c|}{ } & E & 2.0 & Inishi CM & N28E76W & -- & -- & -- \\
\cline{3-9}
\hline
\multicolumn{2}{|c|}{Major} & f' & 2.5 & Inishi CM & N5W81E & -- & -- & -- \\
\cline{3-9}
\multicolumn{2}{|c|}{Discontinuity} & A-1 & 0.7 & Inishi CM & N57W74W & -- & -- & -- \\
\hline
\multicolumn{2}{|c|}{ } & East & -- & Skarn CH & -- & -- & -- & -- \\
\cline{3-9}
\multicolumn{2}{|c|}{Skarn} & \makecell{Mixing \\ with \\ Inishi} & -- & Skarn B & -- & -- & -- & -- \\
\cline{3-9}
\multicolumn{2}{|c|}{ } & West & -- & Skarn B & -- & -- & -- & -- \\
\hline
\end{tabular}
\end{small}
\end{table}

\begin{table}[!ph]
\caption{Final parameter settings for weak layers and geological units in the cylindrical section.
These parameters define the section-specific input set used in the 3D sequential elasto-plastic analyses. 
Strike/dip notation is the same as in Table~\ref{tab:dome_geo}. 
Here, $K_n$ and $K_s$ denote the normal and tangential spring constants of the interface (IF) elements, respectively. 
}
\label{tab:barrel_geo}
\centering
\begin{small}
\begin{tabular}{|c|c|c|c|c|c|c|c|c|}
\hline
\multicolumn{3}{|c|}{Elements of Model} & Width & Rock mass & Strike & \multicolumn{3}{c|}{Interface Element} \\
\cline{7-9}
\multicolumn{3}{|c|}{} & [m] & classification & and dip & Position & $K_{n}$ & $K_{s}$ \\
\multicolumn{3}{|c|}{} & &  & &  & {\scriptsize [GPa/m]} & {\scriptsize [GPa/m]} \\
\hline
\multicolumn{3}{|c|}{Main Rock} & -- & \makecell{Inishi \\ CH-8G} & -- & -- & -- & -- \\
\hline
\multicolumn{2}{|c|}{ }  &   &  & & & dome & 70 & 0.35 \\
\cline{7-9}
\multicolumn{2}{|c|}{ }  &   &  & & & \makecell{cylinder \\ east} & 70 & 0.2 \\
\cline{7-9}
\multicolumn{2}{|c|}{ } & A & 3.5 & \makecell{Inishi \\ CM-4G} & N60W75S & \makecell{cylinder \\ west (near \\ the wall)} & 70 & 0.05 \\
\cline{7-9}
\multicolumn{2}{|c|}{\makecell{Weak Layer \\ \hspace{1cm} \\ \hspace{1cm}}} &   &  & & & \makecell{cylinder \\ west (far \\ side of \\ the wall)} & 70 & 0.1 \\
\cline{3-9}
\multicolumn{2}{|c|}{ } & A upper & 2.5 & Inishi CM & N60W75S & -- & -- & -- \\
\cline{3-9}
\multicolumn{2}{|c|}{ } & H & 2 & \makecell{Inishi \\ CM-6G} & N10E55W & -- & 70 & 0.35 \\
\cline{3-9}
\multicolumn{2}{|c|}{ } & E & 2.0 & Inishi CM & N25E84W & -- & -- & -- \\
\hline
\multicolumn{2}{|c|}{ } &  &  &  & & \makecell{south \\ (1B-10B)} & 70 & 0.2 \\
\cline{7-9}
\multicolumn{2}{|c|}{ } & \makecell{X7 \\ \hspace{0.3cm} \\ \hspace{0.3cm}} & \makecell{0.7 \\ \hspace{0.3cm} \\ \hspace{0.3cm}} & \makecell{Inishi CM \\ \hspace{0.3cm} \\ \hspace{0.3cm}} & \makecell{N10W80E \\ \hspace{0.3cm} \\ \hspace{0.3cm}} & \makecell{north \& \\ south \\ (11B-19B)} & 70 & 0.2 \\
\cline{3-9}
\multicolumn{2}{|c|}{Major} & f' & 0.7 & Inishi CM & N10W80E & -- & -- & -- \\
\cline{3-9}
\multicolumn{2}{|c|}{Discontinuity} & A-1 & 0.7 & Inishi CM & N55W75W & -- & -- & -- \\
\cline{3-9}
\multicolumn{2}{|c|}{ } & I & 0.7 & Inishi CM & N56W63W & -- & -- & -- \\
\cline{3-9}
\multicolumn{2}{|c|}{ } & Cy1 & 0.7 & Inishi CM & N10W80W & -- & -- & -- \\
\cline{3-9}
\multicolumn{2}{|c|}{ } & 4B-15 & 0.7 & Inishi CM & N65E80N & -- & -- & -- \\
\cline{3-9}
\multicolumn{2}{|c|}{ } & 13B-25 & 0.7 & Inishi CM & N0E80E & (6B-11B) & 0.3 & 0.35 \\
\cline{7-9}
\multicolumn{2}{|c|}{ } &  & &  & & (12B-19B) & 0.8 & 0.35 \\
\hline
\multicolumn{2}{|c|}{ } & East & -- & Inishi CH & -- & -- & -- & -- \\
\cline{3-9}
\multicolumn{2}{|c|}{Skarn} & \makecell{Mixing \\ with \\ Inishi} & -- & Inishi CH & -- & -- & -- & -- \\
\cline{3-9}
\multicolumn{2}{|c|}{ } & West & -- & Skarn B & -- & -- & -- & -- \\
\hline
\end{tabular}
\end{small}
\end{table}

\subsection{Dome Section} \label{sec:dome_design}
The dome must withstand the in-situ stress environment associated with $\sim$600~m of overburden and the resulting stress concentration around the excavation; its geometry is therefore critical. 
The initial plan assumed a 14~m rise; based on measured initial stress, a 21~m rise was adopted to achieve a rise-to-span ratio comparable to SK. 
Comparative heterogeneous elasto-plastic analyses for rises of 14~m and 21~m under Cases A and C showed that, under the higher  principal-stress-ratio case (Case C), a 14~m rise produces deep tensile failure reaching $\sim$20~m. Given the maximum anchor length of 22~m and $\sim$4~m embedment required into competent rock, stabilizing the crown would be difficult. With a 21~m rise, tensile failure depth was limited to $\sim$15 m even under Case C. Although this increased excavation volume and prolonged construction, risk under non-nominal stress led to adopting a 21~m rise.

Dome analysis used elasto-plastic sequential simulation: the initial heading was Step 1; each ring was subdivided according to actual support-installation timing, yielding 24 steps to completion (step divisions follow the thick solid lines in Fig.~\ref{fig:division}, including the core boundaries shown in the right panel). 

Although pre-construction investigations indicated competent rock with high shear strength and high elastic modulus, excavation revealed lower elastic moduli than expected. Because a higher elastic modulus often correlates with higher strength, this raised concern about the appropriateness of the strength settings. The elastic modulus of 8~GPa assigned to aplite/CH-class bedrock in the dome is low for CH. Until reducing the elastic modulus while maintaining strength was validated, support was conservatively implemented. Additionally, when rocks of different elastic moduli contact in a continuum model, failure may initiate in the stiffer unit due to stiffness contrast, complicating rational design. Countermeasures included: (i) introducing interface (IF) elements~\cite{IF} along major weak layers controlling deformation, allowing discontinuous slip/opening; and (ii) maintaining strength consistent with geological evaluation while applying averaged equivalent elastic modulus for stiffness.

In the analytical geological model, IF elements were introduced to reproduce opening/slip along major weak layers for which solid-element modeling would be inappropriate. In the dome, IF elements were assigned to the lower surface of Weak Layer A and upper surface of Weak Layer H, positioned per borehole logs indicating fracture cores relative to reduced-quality zones. 
IF properties were 
determined through parameter studies by adjusting the normal and tangential spring constants while keeping the rock-mass elastic modulus fixed, to reproduce the measured displacements, resulting in
$K_n$ = 70~GPa/m and $K_s$ = 0.7~GPa/m. 
After introducing IF elements to mitigate stiffness-contrast effects, the stiffness of Weak Layers A and H, including zones of reduced rock-mass quality associated with discontinuities, was set to 4~GPa and 6~GPa, respectively, based on parameter studies using measured displacements. The final parameters for weak layers and geological units in the dome section are summarized in Table~\ref{tab:dome_geo}.

The primary objective of dome support is to prevent collapse of loosened zones near the crown. PS anchors and shotcrete were designed for long-term stability. Following power-station practice, uniform-thickness blocks were assumed for tensile-failure zones, and wedge-shaped blocks were defined for shear-failure zones by connecting the most failure-prone planes under circumferential compression. Support quantities, including PS anchors, were computed using limit-equilibrium analyses. To account for uncertainties in discontinuity positions and rock-mass modeling, support layout was derived from failure-zone distributions predicted by excavation analysis, using the maximum failure depth within a 2.5~m radius on the wall as the design criterion. 
The design concepts are as follows
 (see Table~\ref{tab:app_sup_design_dome} for schematic illustrations):
\begin{description}
\item[Suspension capacity: ] Each PS anchor must support the weight of the rock block, using pretension and shotcrete shear resistance.
\item[Prevention of dislodgement: ] Rock between anchors must be safely supported by shotcrete shear capacity.
\item[Tensile-failure block collapse: ]  Even if the tensile plastic (loosened) zone collapses as a single block, suspension by PS anchors plus peripheral shotcrete shear must suffice. In the design, the block is modeled as a uniform-thickness block extending through the depth of the plastic (loosened) zone. Because shotcrete shear resistance can only be mobilized at the periphery, this condition is more stringent than simple suspension.
\item[Shear-failure wedge collapse: ]  For shear zones due to circumferential uniaxialization, stability was verified for wedges formed along mobilized joint planes determined by friction angle 
$\Phi$. 
Plastic (loosened) zone depth came from analysis; 
$\Phi$
used for mobilized planes corresponded to the peak bedrock strength (aplite-equivalent). 
\end{description}

Because the central crown becomes increasingly difficult to access, and for safety against block collapse, many PS anchors were installed early, in addition to those required strictly for suspension. Wherever possible, longer anchors were used to stitch dominant weak layers. The design shotcrete shear strength was initially 0.5~MPa; as behavior indicated reduced block-formation risk, from mid-Stage 4R-1 (diameter $>$ 40~m), the design shear strength was revised to 3.0~MPa (one-sixth of nominal shotcrete compressive strength, 18~MPa), consistent with quality control test results, data variability, and past performance. Final dome support quantities are summarized in Table~\ref{tab:dome_support}.

\begin{table}
\caption{Summary of dome support quantities.} \label{tab:dome_support}
\begin{small}
\begin{tabular}{|c|c|c|c|c|c|c|}
\multicolumn{7}{l}{Prestressed anchors} \\ 
\hline
Spec. & Quantity & Length & \makecell{Total length \\ of anchors} &  \makecell{Initial tension \\ of anchor} & Yield load & \makecell{Shear \\ capacity} \\
      &  & [m] & [m] & [kN] & [kN] & [kN] \\
\hline
300~kN & 317 & 8, 9, 10, 12 & 2,807 & 240--300(\$) & 444 & -- \\
600~kN & 262 & 9, 10, 12, 14, 16, 22 & 3,355 & 520--600(\$) & 888 & -- \\
\hline
Total  & 579 & --  & 6,162 & 247,260 & -- & -- \\
\hline
\multicolumn{6}{l}{(\$) Initial tension of each anchor} \\ 
\multicolumn{6}{l}{\hspace*{0.5cm}}
\end{tabular}
\begin{tabular}{|c|c|c|c|}
\multicolumn{4}{l}{Shotcrete} \\ 
\hline
Category & Thickness & \makecell{Shear \\ capacity
(\#)
} & \makecell{Fiber \\ reinforcement} \\
     & [cm] & [kN/m] & \\
\hline
Primary     & 15   & 450 & None \\
Secondary   & 15   & 450 & Mix \\
\hline
Total   & 30 & 900 & -- \\
\hline
\end{tabular}\\
\begin{tabular}{cl}
(\#)
 & 
Nominal strength ($\sigma_{28}$): 18N/mm$^2$ = 18~MPa.\\
\end{tabular}
\end{small}
\end{table}

\subsection{Cylindrical Section} \label{sec:barrel_design}
Cylinder analysis also used elasto-plastic sequential simulation. Excavation of the spiral working drift (replacing Approach Tunnel No. 3) and benches 1B-19B were defined as steps, yielding 44 steps (including dome steps). 

In the dome, aplite was present; bedrock was defined as a mix of aplite and CH-class rock (CH-1 in Fig.~\ref{fig:cavern_geo}). In the cylinder, aplite was scarce; bedrock was defined as CH class (CH-B in Fig.~\ref{fig:cavern_geo}). The elastic modulus was set at 8 GPa, consistent with the dome. During cylinder excavation, Weak Layer X7 (minor influence in the dome) was added to the model as a highly continuous plane causing significant displacement. 
To reproduce behavior in the analytical geological model, IF spring constants for Weak Layers A and X7 were revised during excavation; each revision required recalculation of loosened zones and support adjustments. Additional discontinuities---such as inclined planes functioning as potential slip surfaces and highly continuous clay-bearing planes identified during cylinder excavation---were incorporated into the geological model and updated as excavation progressed. Their positions, strikes, and dips were revised accordingly in the analysis model definitions. The final parameters for weak layers and geological units in the cylindrical section are summarized in Table~\ref{tab:barrel_geo} 
(see the lower panel of Fig.~\ref{fig:cavern_geo}).

\begin{table}
\caption{Summary of cylinder support quantities. 
} \label{tab:barrel_support}
\begin{small}
\begin{tabular}{|c|c|c|c|c|c|c|}
\multicolumn{7}{l}{Prestressed anchors} \\ 
\hline
Spec. & Quantity & Length & \makecell{Total length \\ of anchors} & \makecell{Initial tension \\ of anchor} & Yield load & \makecell{Shear \\ capacity} \\
      &  & [m] & [m] & [kN] & [kN] & [kN] \\
\hline
300~kN & 1,215 & 8, 9, 10, 12, 14 & 13,040 & 200--300(\$) & 444 & 301(*) \\
600~kN & 305 & 8, 9, 10, 12, 14 & 4,070 & 200--600(\$) & 888 & 301(*) \\
\hline
Total  & 1,520 & --  & 17,110 & 369,640 & -- & -- \\
\hline
\multicolumn{7}{l}{(\$) Initial tension of each anchor  (*) Shear capacity was counted at priority monitoring areas only} \\ 
\multicolumn{6}{l}{\hspace*{0.5cm}}
\end{tabular}
\begin{tabular}{|c|c|c|c|}
\multicolumn{4}{l}{Shotcrete (at general area)} \\ 
\hline
Category & Thickness & \makecell{Shear \\ capacity
(\#)
} & \makecell{Fiber \\ reinforce- \\ ment} \\ 
     & [cm] & [kN/m] & \\
\hline
Primary     & 20   & 600 & None \\
Secondary   & 10   & 300 & None \\
\hline
Total   & 30 & 900 & -- \\
\hline
\end{tabular}
\begin{tabular}{|c|c|c|c|}
\multicolumn{4}{l}{Shotcrete (at priority monitoring areas)} \\
\hline
Category & Thickness & \makecell{Shear \\ capacity
(\#)
} & \makecell{Fiber \\ reinforce- \\ ment} \\
     & [cm] & [kN/m] & \\
\hline
Primary     & 20   & 100(*) & None \\ 
Secondary   & 10   &  50(*) & None \\ 
\hline
Total   & 30 & 150(*) & -- \\
\hline
\end{tabular}
\begin{tabular}{cl}
(*) & In stability checks for countermeasure design at the priority monitoring areas, the shear \\
 & capacity of shotcrete was not counted (conservative assumption), except where explicitly \\
& noted in Sec.\ref{sec:crack}.\\
(\#)
 & 
Nominal strength ($\sigma_{28}$): 18N/mm$^2$ = 18~MPa.\\
\end{tabular}
\end{small}
\end{table}

During excavation, comparison of west--east deformation behavior showed that Case D (Table~\ref{tab:init_stress}), with higher lateral stress, matched observed displacement patterns more closely. Because greater lateral stress enlarges shear-failure zones and is conservative for support design, the cylindrical-section analysis adopted Case D', obtained by dividing each principal stress of Case D by 1.2, reflecting overburden conditions and observed behavior. 

The primary objective of cylinder support is to prevent wedge sliding along the cavern wall. Arbitrary sliding wedges may form within loosened zones; all patterns must satisfy the required safety factor. For plastic (loosened) zones at completion, PS-anchor number and specifications were determined by limit-equilibrium analysis. For long-term support design, shotcrete thickness was set to the finished design value of 30~cm. Consistent with power-station practice, rock-bolt restraint was not considered in global long-term stability. 

While elasto-plastic analysis holds shear zones at peak strength (bearing stress), support design used residual friction angles for sliding planes with cohesion set to zero. 
The circumferential width of a 3D sliding wedge was defined as the azimuthal span obtained from the stereographic wedge curve, scaled to the plastic zone depth 
(see Fig.~\ref{fig:app_3d_block} for the definition of the wedge width in the 3D treatment). 
Unlike earlier 2D practices that considered only shotcrete restraint on the wedge bottom per unit depth, the 3D treatment allows restraint on vertical faces to be included, enabling more realistic, rational design. Shotcrete shear strength was set to 3.0~MPa, consistent with the final dome design. 
Final cylinder support quantities are summarized in Table~\ref{tab:barrel_support} (priority monitoring areas in Table~\ref{tab:barrel_support} are introduced and discussed in Sec.~\ref{sec:info}).

\section{Information-Based (Observational) Design and Construction Approach} \label{sec:info} 
Even with substantial pre-excavation investigations, fully characterizing the surrounding rock mass was impractical. Ensuring stability solely by pre-construction design was therefore infeasible; an information-based design and construction approach---incorporating excavation-induced behavior measurements and geological observations---was essential~\cite{info_apply}. 
This information-based approach is an integral part of the design philosophy for the HK cavern. 

\subsection{Rock-Mass Behavior Monitoring} \label{sec:info_mon}
Behavior was monitored primarily via extensometers and PS-anchor (prestressed-anchor) load cells
to verify predicted deformation and assess the performance of the support elements (shotcrete and PS anchors) during excavation. 
Multi-stage extensometers (maximum length 25~m; six fixed points) measured displacement across multiple intervals. 
Typically, the deepest point (25~m) served as reference; nominal intervals were 0, 2.5, 5, 10, and 15~m, with displacement derived from five steel wires and potentiometers (intervals adjustable;
see Fig.~\ref{fig:app_extensometer} for schematic diagrams and photographs). 
PS-anchor load cells measured head stress by strain gauges; load increments from initial tension were converted to strand elongation using elastic modulus and cross-section
(see Fig.~\ref{fig:app_load_cell} for schematic diagrams and a photograph).
Shotcrete stress gauges were installed at the outermost dome ring (6R) to address concerns about longitudinal uniaxialization during cylinder excavation. 

Dome instrument layout (Fig.~\ref{fig:cavern_wl}, upper; 
see also Fig.~\ref{fig:app_bird_inst_wl} for a bird's-eye view): 
given $\sigma_1$ approximately east--west, main measurement cross-sections were oriented east--west and north--south, with auxiliary sections at 45$^\circ$. Extensometers were placed at 1R (center and N-S), 2R (E-W and N-S), and 3R-5R (eight directions). 
None were installed at 6R due to proximity to 1B. 
Load cells were co-located near extensometers at 1R; at 2R-3R, at 45$^\circ$ intervals; and from 4R onward, at $\sim$10~m spacing including 6R, yielding 16 directions. 
After early confirmation of correspondence between displacement increments and load-cell readings (converted to displacement), load cells were omitted where extensometers were installed from 4R onward. 
At 6R, eight shotcrete stress gauges were added. Additional instruments were installed as needed during the information-based process. Fiber-optic displacement meters~\cite{fiber_displacement_sensor} were installed as part of R\&D, confirming behavior dominated by fracture opening and slip
(see Fig.~\ref{fig:app_fiber_sensor} for schematic diagrams and a photograph).
Instruments were installed after local excavation; excavation-induced displacements during advance were therefore not captured directly and were estimated to be $\sim$50\% of the total excavation-induced displacement (from the pre-excavation state to completion) with location-dependent variation 
(see Fig.~\ref{fig:division} for the overall excavation and support sequence; for the dome section in more detail, see Fig.~\ref{fig:app_excavation_procedure}).

The cylinder instrument layout used eight directions at 45$^\circ$ intervals. Extensometers were installed at benches 1B, 5B, 9B, and 13B in all eight directions. 
Load cells were generally installed between extensometers (eight per bench). 
Additional devices were added for weak-layer movement, including fiber-optic meters. At bench 17B (originally planned to be omitted), four extensometers were installed (E, W, N, S) to monitor creep after completion. 
At bench 19B, one extensometer monitored post-excavation movement of Weak Layer A. 
Final layouts are in Fig.~\ref{fig:cavern_wl} (lower; see also Fig.~\ref{fig:app_bird_inst_wl} for a bird's-eye view). Depth profiles along main measurement lines, encompassing both the dome and cylindrical sections, are shown in Figs.~\ref{fig:supportNS} (N-S) and \ref{fig:supportEW} (E-W), prepared from data after completion.

\subsection{Geological Observations} \label{sec:info_geo}
\begin{figure}[htbp!]
\begin{center}
\hspace*{-1.0cm}
\includegraphics[width=1.2\linewidth]{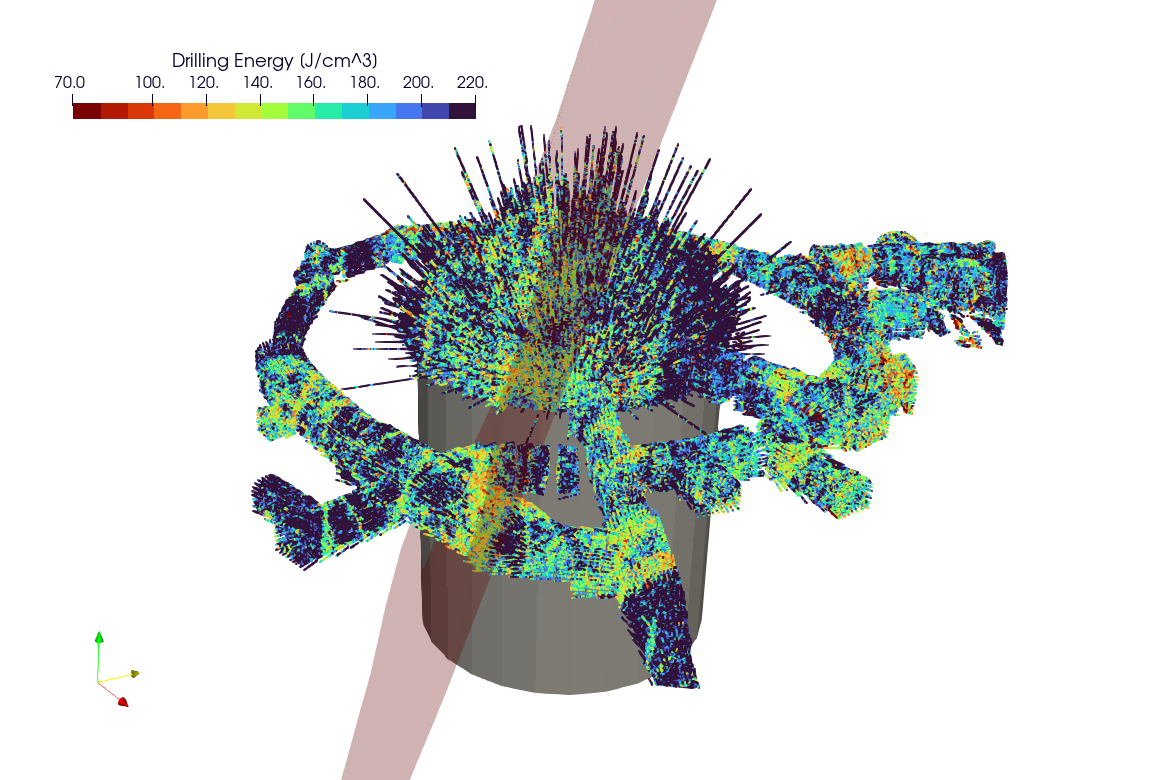}
\caption{
Bird's-eye view of drilling energy derived from Measurement-While-Drilling (MWD) data during blast-hole and PS-anchor/rock-bolt drilling, showing a distinct low-energy zone along Weak Layer A (Geographia estimate). 
}
\label{fig:weaklayer_top}
\end{center}
\end{figure}

\begin{figure}[htbp!]
\begin{center}
\includegraphics[width=1.0\linewidth]{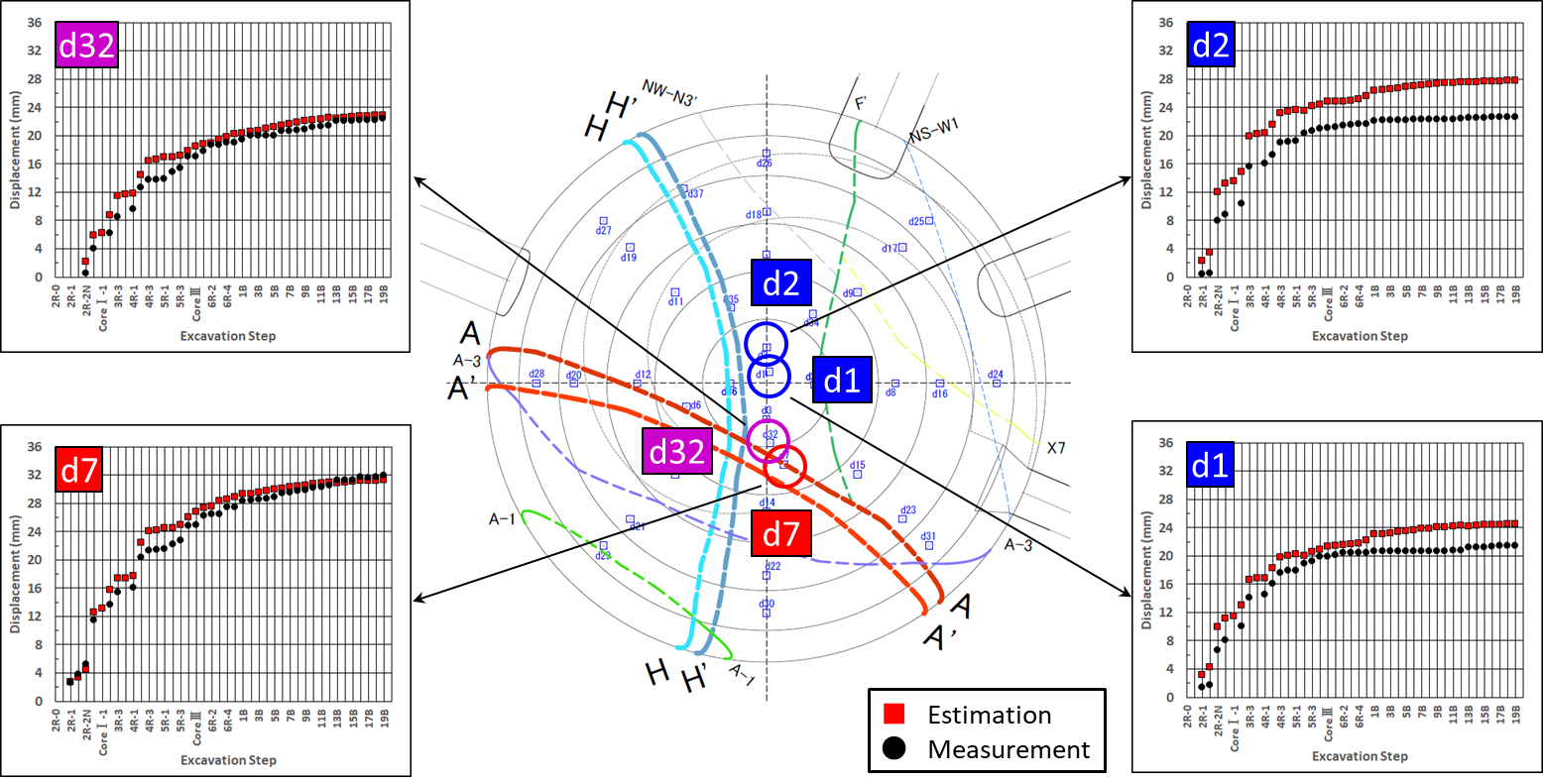}
\caption{
Positions of extensometers along Weak Layers A and H in the dome and step-dependent wall displacement. 
Across the four profiles, the final model reproduces the observed displacement trends. 
Two are consistent overall (with minor exceedances attributable to creep-like behavior; see Sec.~\ref{sec:creep}), while the other two are conservatively overpredicted, supporting the observational design framework.
}
\label{fig:weaklayer_bottom}
\end{center}
\end{figure}

Geological observation included wall mapping, use of Measurement-While-Drilling (MWD) data, and BTV surveys. Wall mapping provided characteristics, strike/dip, and continuity of weak layers, and rock-mass classification. From MWD data (e.g. drilling speed, impact pressure), drilling energy was computed~\cite{MWD}; it correlated well with rock-mass classification, and was used with mapping to estimate weak-layer strike/dip and identify relatively weakened zones. 
Although blast-hole MWD reflects excavated regions, continuity with walls made it useful for continuity evaluation; 
MWD from PS-anchor and measurement boreholes characterized the rock behind the wall requiring stabilization, and verified PS-anchor embedment. Fig.~\ref{fig:weaklayer_top} shows Weak Layer A conditions from MWD,
highlighting a distinct low-energy zone.
Fig.~\ref{fig:weaklayer_bottom} shows the positions of extensometers along Weak Layers A and H in the dome and the step-dependent wall displacement.
Comparison with the model confirmed that actual behavior was captured and that displacements during cylinder excavation were small and convergent.

BTV provided information on weak layers and discontinuities in the deep crown region and deep sidewalls, aiding root-cause identification when discrepancies arose between predictions and measurements. 
Consistency between geological evaluation and the analysis model was continuously discussed, and model revisions were made as needed (e.g., weakened zones above intersections of A and H). Data were consolidated in Geographia~\cite{Geographia}; updates to continuity, strike, and dip improved accuracy and supported interpretation and model updates. Final results are in Fig.~\ref{fig:cavern_geo}.


\subsection{Management Criteria} \label{sec:management}
To ensure safe implementation of the information-based approach, three management criteria were defined:

\begin{description}
\item[Criterion I:] ``Cause/countermeasure review level.''
\item[Criterion II:] ``Countermeasure implementation level (including potential work suspension).''
\item[Criterion III:] ``Work-suspension level.''
\end{description}

Specific actions were prescribed for exceedance of each threshold. Criterion I was set based on predicted displacement (or load). 
Criterion II corresponded to 90\% of the PS-anchor steel yield load; for extensometers, thresholds were recalculated using the yield loads of surrounding PS anchors. 
Criterion III was set at 95\% of the PS-anchor yield load. Because support-element integrity is essential for stability, Criteria II and III were based on PS-anchor yield load.

Operationally, anticipated exceedance of Criterion I triggered model review and updates to displacement predictions and support design, minimizing schedule impact. Upon reaching Criterion II, decisions focused on whether displacement acceleration exhibited convergence or divergence. If convergent, sufficient margin remained before Criterion III, allowing technical review and potential model updates without immediate countermeasures or suspension.

\subsection{Issues and Countermeasures} \label{sec:issues}
\subsubsection{PS-Anchor Overloading (Dome and Cylinder)} \label{sec:overload}
Early in the dome excavation, the deformation along weak layers exceeded initial predictions. The model was updated by reducing bedrock elastic modulus from 16~GPa to 8~GPa, incorporating newly identified reduced-quality zones, and introducing IF (interface) elements for major weak layers. The updated model matched (or bracketed) observed behavior and provided the basis for redesigning supports to stabilize predicted plastic (loosened) zones; additional support was installed where deficiencies were identified. 

At the completion of Stage 2R, load cells recorded unexpectedly large increases on PS anchors along Weak Layer A. If anchors exceed their capacity, rupture may occur, potentially triggering overload propagation and a catastrophic roof collapse; ensuring anchor integrity is therefore fundamental. The loads were checked by jacking the adjacent anchors; nineteen anchors at risk of overloading before completion were scheduled for unloading. 
To prevent unintended load transfer during the unloading operations, additional anchors were installed and preloaded in advance, in accordance with established practice. 
Analysis suggested that the deformation along weak layers was underestimated. No significant load increases were observed on adjacent anchors during unloading. Subsequently, the deformation allowances were computed for each anchor location based on predicted completion displacements and past performance, and anchor loads were managed accordingly. Newly installed anchors received lower initial pretension as a precaution; however, their support contribution was correspondingly reduced in limit-equilibrium checks. 

In the cylinder, despite model updates informed by dome experience, unexpected weak-layer movement led to overloads along Weak Layers A and X7 at bench 4B completion. Countermeasures included installing additional anchors and performing unloading. Where loads exceeded $\sim$400~kN and unloading was infeasible, strands were cut and the reduction compensated by additional anchors. For later overloads, protective nets were installed to prevent head projection, and additional support was implemented without cutting/unloading. Review of past performance showed that anchors exceeding capacity retained 
deformation margins of roughly 10-fold 
before rupture, and even in rupture events, head-projection risk was extremely low. 

Discontinuities acting as slip surfaces at the wall, though  not laterally extensive, could not always be identified during bench-wall inspections and occasionally caused localized large displacement. 
To mitigate this risk, nominal PS anchors installed from benches 7B--17B were designed as 12~m anchors with an 8~m free length. Exceedances of predicted displacements and loads during cylinder excavation prompted model updates; large displacements along weak layers on the west side led to revisiting initial stress. As benches advanced and the distance from the face increased, continued weak-layer movement caused absolute displacements to exceed predictions, necessitating repeated revisions of IF spring constants throughout the process.

\subsubsection{Dome Section: Massive Key Block}
In the dome, the main weak layers include Weak Layer A (south side; strike NW-SE; dip SW), Weak Layer H (west side; strike N-S; dip west), and Weak Layer f' (east side; strike N-S; dip east) (Fig. 5, upper). The area enclosed by these layers could form a cantilevered block. If a discontinuity with NW-SE strike and NE dip existed on the north side, a massive key block~\cite{KeyBlock} might form. Although numerous NW-SE fractures were observed on the northeast side, none showed high continuity in the spiral pilot heading; the risk was deemed low but not negligible. Stabilizing such a block would require $>$ 200 additional PS anchors. 

Before excavating the northern part of ring 3R, four extensometers were installed along Weak Layers H and f' to directly monitor possible key-block movement. 
 Devices were positioned to straddle weak layers identified by BTV, enabling direct block-movement monitoring (instruments d33--d36 in Fig.~\ref{fig:cavern_wl}, upper; see also Fig.~\ref{fig:app_dome_keyblock} for a bird's-eye view). As 3R progressed counterclockwise, post-blast behavior near the critical weak layers was monitored to evaluate the key-block formation. 
Cantilever-beam calculations indicated large deflections if such a block formed (58~mm at 3R completion; 4,820~mm at 4R completion). Observations showed 
a decrease in the risk 
through 3R and 4R. Ultimately, no movements suggestive of a massive key block were observed through dome completion, and large-scale additional support was unnecessary.

\subsubsection{Dome Section: Tensile-Failure Rock Blocks}
The greatest demand for PS-anchor suspension capacity in the dome arises from tensile-failure blocks. During rings 1R-3R, anchors were installed with fixed ends above weak layers and, where possible, stitching across them. This mitigated future key-block risk, ensured crown stability, and minimized rework. Up to 3R completion, observed displacements continued to exceed predictions, requiring ongoing model updates and conservative support. As excavation progressed through 4R--5R, the dome's 3D arch effect increased, and consistency between analysis and observations improved.

Throughout final dome stages, no behavior indicative of tensile-failure block occurrence was observed. Safety was ultimately ensured by installing sufficient anchors to provide adequate suspension capacity for potential blocks formed by contiguous tensile plastic (loosened) zones, while accounting for shotcrete shear resistance of 3~MPa.

\subsubsection{Cylinder Section: Shotcrete Cracking} \label{sec:crack}
\begin{figure}[bthp!]
\begin{center}
\includegraphics[width=0.75\linewidth]{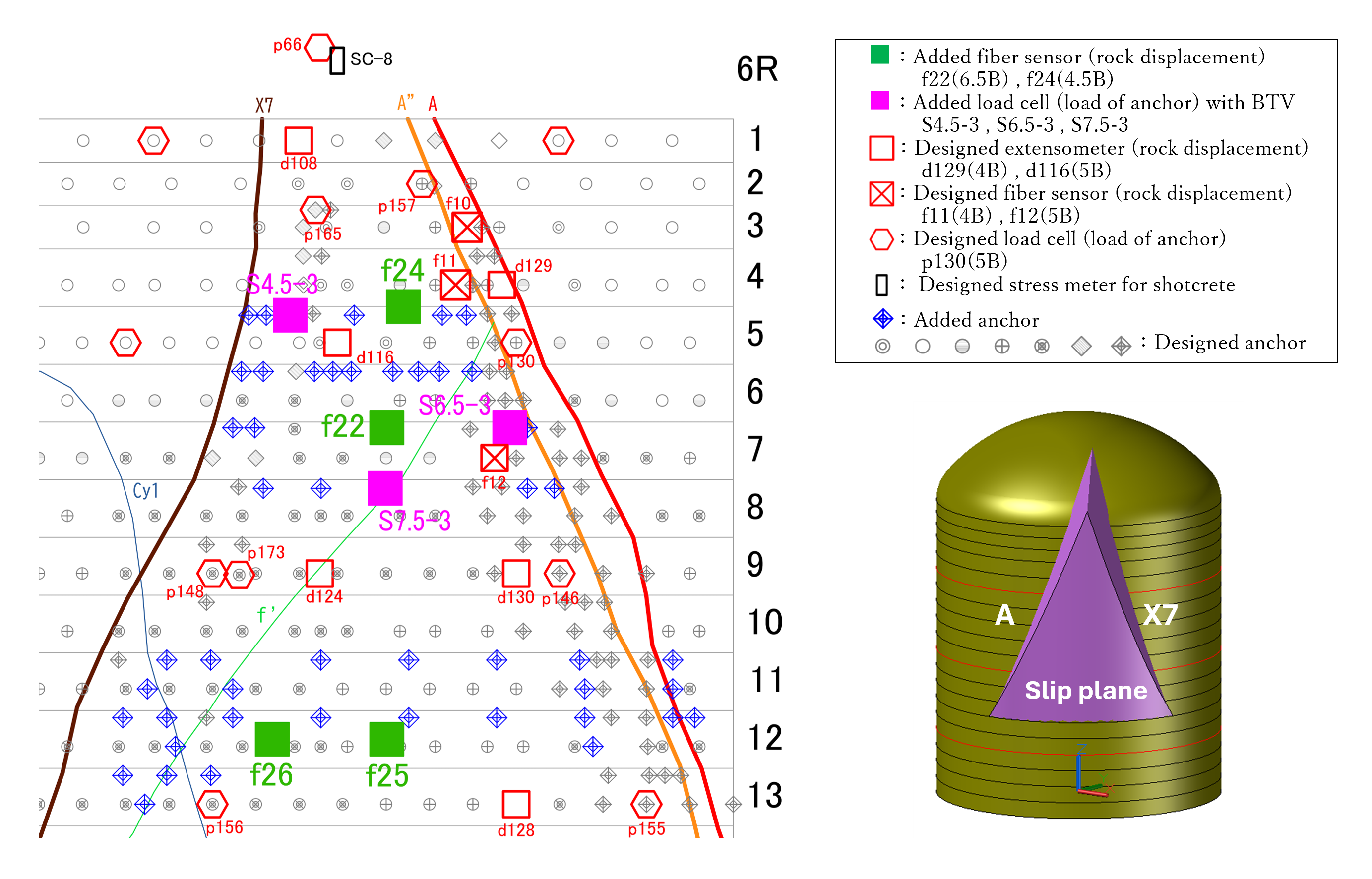}
\includegraphics[width=0.75\linewidth]{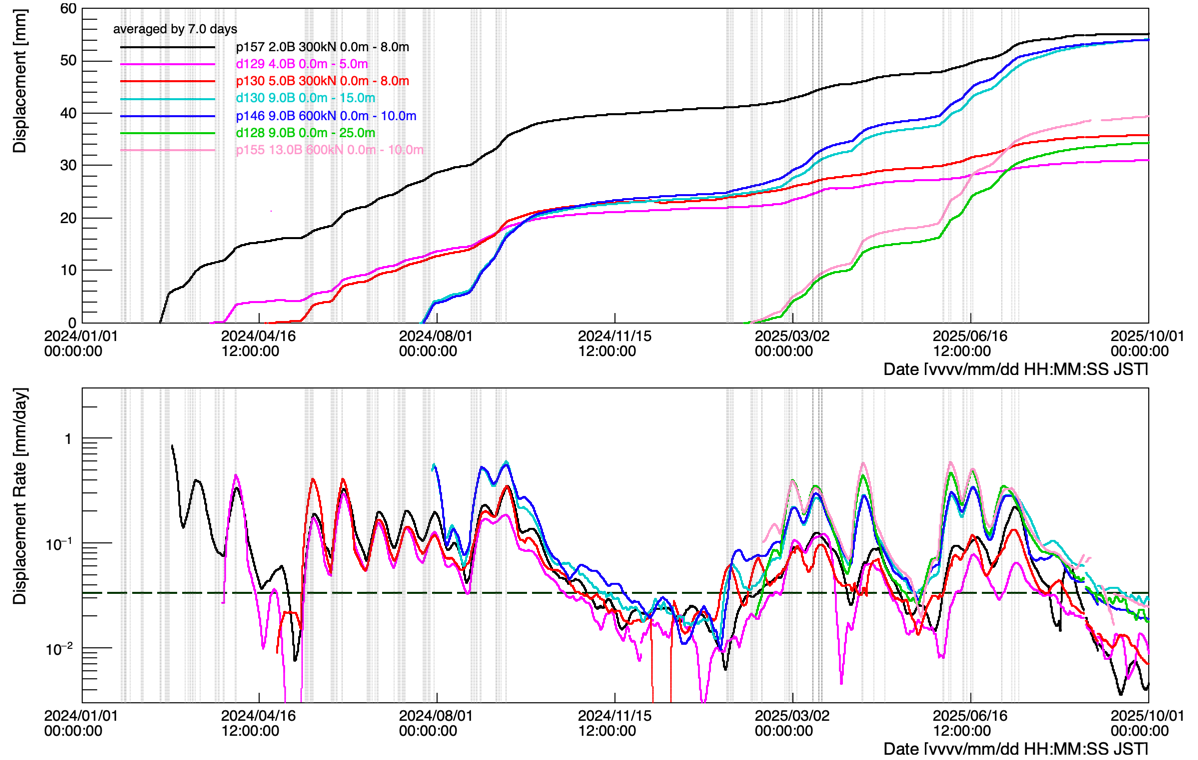}
\caption{
(Upper left) Wall development diagram (southeast) showing added PS anchors and instruments. (Upper right) Assumed sliding block in the southeast. (Bottom) Displacement and displacement rate of instruments crossing Weak Layer A (southeast). The legend for each colored line is indicated in the figure. Load-cell readings are converted to strand elongation using steel modulus. 
The gray vertical dotted lines indicate the blasting timing at the outermost perimeter, which has significant impacts on the displacement. 
The long-dashed line in the displacement rate panel indicates a creep rate of 1~mm/month.
Displacements increase primarily after blasting. During non-excavation periods, the displacement rate decays, indicating converging behavior. This convergence continues after excavation completion, providing evidence of cavern stabilization. 
}
\label{fig:block}
\end{center}
\end{figure}

After completing outer excavation at bench 12B (12 September 2024), the vertical stress measured by shotcrete stress gauge sc-8 (see the upper left panel of Fig.~\ref{fig:block}) in the southeastern dome (6R) decreased to $\sim$0 from a peak of $\sim$32~MPa in late July. Inspection revealed surface cracks in shotcrete. While repair and core blasting at 13B continued, on 27 September 2024, vertical delamination cracks were discovered in the southeast along Weak Layer X7 between benches 4B and 6B 
(see Appendix~\ref{sec:app_shotcrete_cracks} for the excavation status at the onset of shotcrete cracking). 

In the cylinder, the convergence of displacement with bench advance at instruments crossing major weak layers was slower than expected, and displacements exceeding predictions persisted---ultimately leading to shotcrete cracking. 
Notably, cracks occurred even after slow convergence with excavation advanced more than six benches below the affected area. Excavation was temporarily suspended and countermeasures discussed. 
It was determined that a large sliding block could form within the wedge bounded by Weak Layers A and X7 in the southeast, and on the footwall side of Weak Layer A on the west. 
Scaffolding was erected from the face, and additional PS anchors were installed in previously excavated areas; safety measures against shotcrete spalling were implemented in parallel, and a monitoring plan was established. Details follow.

\paragraph{\bf Support Design for Priority Monitoring Areas} 
After the detection of 
cracks along Weak Layer X7, drone-based inspection surveyed shotcrete cracks throughout the cylinder, revealing additional cracks along Weak Layers A-1 and A. Concerns arose that shotcrete resistance---critical for stability of arbitrary sliding blocks---might underperform. 
The southeast 
wall 
bounded by Weak Layers X7 and A and the west 
wall 
near Weak Layer A showed significant displacement, raising concerns about crack propagation. These areas were designated priority monitoring areas for individual design review. 
Fig.~\ref{fig:block} (bottom) shows time-dependent displacement and rate for instruments crossing Weak Layer A in the southeast. 

For general areas, design conditions were revised: assuming the possibility of losing resistance on one vertical shotcrete face due to cracking, stability checks considered resistance from only one vertical face, and additional support was installed where needed. In practice, cracks along Weak Layer A-1 were treated under the general-area criteria, and no additional support was required in the general areas. Priority monitoring areas were therefore limited to the southeast wedge bounded by Weak Layers A and X7 and the west-side area near Weak Layer A.

For the southeastern priority monitoring area, design assumed slip surfaces could form anywhere within the plastic (loosened) zone of the wedge bounded by Weak Layers X7 and A. A massive block closing at bench 13B, with maximum plastic depth of $\sim$8~m, was considered; stability was verified for this and for closures at higher benches. 
The shear resistance of PS anchors were included based on a past example, differing from the general-area support design.
In view of cracks observed along bench boundaries near intersections with major weak layers, the shear resistance of shotcrete on horizontal surfaces was excluded from consideration as a conservative measure. Larger blocks were not considered beyond 13B because Weak Layer X7 exits outside the wall there. Fig.~\ref{fig:block} (upper left) shows PS-anchor reinforcement and monitoring positions; Fig.~\ref{fig:block} (upper right) shows the assumed southeastern block. 

On the west side, although no weak layer corresponding to Weak Layer A was identified, significant wall displacement on 
the footwall of Weak Layer A 
prompted the consideration of a hypothetical northern discontinuity. 
A block with plastic depth of $\sim$13~m at 4B was assumed for priority monitoring area stability. During reinforcement, discontinuities contributing to wedges and potential slip surfaces were identified on the north side, leading to model updates (features 4B-15 and 13B-25 in Table~\ref{tab:barrel_geo}). As excavation progressed, larger western blocks had to be considered. For blocks extending beyond 14B, horizontal shotcrete shear resistance was assumed to be 0.5~MPa (based on performance at similar sites), including PS-anchor shear resistance.

For general sections, resistance considered in sliding-block checks included one vertical shotcrete face, horizontal faces, shotcrete shear of 3~MPa, and PS-anchor clamping/anchoring. In priority monitoring areas, shotcrete shear resistance was neglected; PS anchors were assumed to provide clamping, anchoring, and shear resistance.

Regarding assumed slip surfaces, cohesion was expected to be present based on the 
considerations listed below; 
however, because it was not included in the initial HK design guidelines and had no precedent in large underground caverns, cohesion was conservatively set to zero. Once excavation progressed and cracks appeared, revising this assumption was judged impractical, as introducing parameters not foreseen in the original design philosophy after onset of instability could compromise safety. 
\begin{itemize}
\item In preliminary design, cohesion was set to $C_r = 0$, 
with the anticipation of 
highly continuous fractures conceivably appearing in plastic (loosened) zones.
\item Face observations and BTV surveys in the cylinder confirmed 
that there was 
no extensive single discontinuity forming a slip surface.
\item If multiple fractures were assumed to connect into a slip surface, newly connected segments would likely exhibit shear strength close to 
that of 
intact rock, implying non-zero cohesion.
\item Even within plastic (loosened) zones, deformation tends to localize along existing fractures; without highly continuous fractures, the likelihood of
the formation of 
a smooth, large slip surface is low.
\end{itemize}

\paragraph{\bf Countermeasures}
Stability calculations required installing $\sim$50 additional anchors in the southeast and $\sim$40 in the west. To ensure stability for blocks closing at higher levels, additional anchors were needed from bench 4.5B (between 4B and 5B) in the southeast and from 6.5B in the west. Five-meter-wide scaffolding was installed from the face; anchors were drilled using skid-mounted rigs. To bring heavy equipment into the cylinder, the spiral working drift was restored
(see Appendix~\ref{sec:app_shotcrete_cracks}); 
a rough terrain crane and other equipment were transported via Approach Tunnel No. 4. In total, 24 anchors were installed in the southeast and 22 in the west using scaffolding; additional monitoring instruments were installed simultaneously. These measures took until 20 January 2025; excavation of bench 13B resumed on 21 January. 

\paragraph{\bf Shotcrete-Crack Safety Measures} 
In cracked areas, rope-access crews removed potentially detached shotcrete, conducted crack inspections and sounding tests to identify hollows, and marked critical points. To ensure safety for work directly below crack zones, rockfall-protection nets were installed to prevent free-falling shotcrete fragments; nets were also installed on the northeast side of Weak Layer X7 before excavation completion.

\paragraph{\bf Monitoring} 
After resumption, monitoring in priority monitoring areas focused on signs of sliding. New instruments tracked potential movement along assumed slip surfaces between major weak layers and the wall. 
Conditions for continuing work were set as follows~\cite{Fukui}: 
\begin{enumerate}
\item After each blast, readings had to show convergence within $\sim$2~hours (5-min sampling) before resuming work inside the cylinder.
\item One week after blasting, the five-day average displacement velocity along assumed slip surfaces had to remain $<$0.1~mm/day before proceeding to the next bench.
\end{enumerate}
The second condition reflects observations during suspension periods, where displacement increases in major weak layers exceeded 0.1~mm/day; if an assumed slip surface formed and reached behavior similar to major weak layers, creep would be expected. Multiple instruments monitored assumed slip surfaces; attention was paid to coherent behavior across devices, indicating potential surface formation.

Drone-based crack inspections were conducted: after each blast, the upper surface of the blasted area was monitored; full scans were performed weekly. 3D scans of cylinder walls and displacement-contour maps were useful for crack detection; after installing protection nets, 3D scanning in priority monitoring areas became difficult.

During excavation at 16B, cracks were discovered along Weak Layer X7 in the northeast; additional chipping and nets were installed. Monitoring continued, including the northeast, and excavation proceeded cautiously. On 31 July 2025, excavation---including support installation at 19B---was completed. 

\subsubsection{Creep Displacement} \label{sec:creep}
Even after excavation was completed, continuous time-dependent displacements (``creep'') were recorded, primarily by extensometers intersecting the major weak layers (see Fig.~\ref{fig:block}, bottom). For a representative instrument intersecting Weak Layer A (extensometer d130), the displacement rate dropped below 1~mm per month roughly two months after completion and has remained below this level, indicating a convergent trend. At that low rate, a cumulative displacement of 5.1~mm was recorded over the five-month period from 1 August to 31 December 2025.

\begin{figure}[bthp!]
\begin{center}
\includegraphics[width=1.0\linewidth]{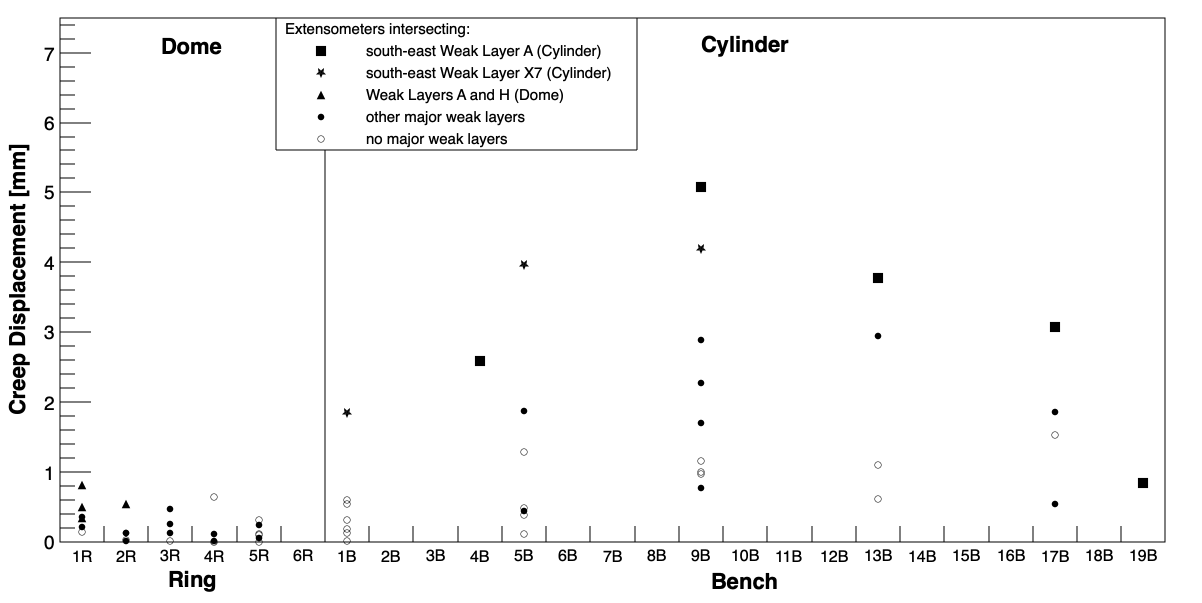}
\caption{
Cumulative creep displacements measured by all extensometers installed in the dome and cylinder over the five-month period from 1 August to 31 December 2025. The horizontal axis indicates each extensometer location (ring number in the dome and bench number in the cylinder). Symbol types indicate extensometer categories (see legend), including whether an extensometer intersects a major weak layer.
}
\label{fig:creep}
\end{center}
\end{figure}

Fig.~\ref{fig:creep} summarizes the cumulative creep displacements measured by all extensometers installed in the dome and cylinder over the same period. Larger creep displacements are observed for extensometers intersecting the major weak layers and for locations near the central region of the cylinder. Monitoring continues with d130 and the other instruments.

\section{Conclusions} \label{sec:conclusion}
The excavation of the Hyper-Kamiokande cavern 
is complete. 
A vertically oriented, dome-capped cylindrical design was chosen to optimize cost and performance. 
Its successful completion represents a major milestone for HK itself and a significant achievement in rock engineering~\cite{WatanabePrize}, 
and is summarized in the following:

\subsubsection*{(i) Incorporated methodology}
\noindent
The project established and demonstrated an integrated, information-based (observational) design and construction framework, characterized by the following elements:
\begin{itemize}
\item A systematic workflow was implemented to integrate behavior measurements and geological observations into successive design updates and operational decisions (Sec.~\ref{sec:info_mon}, Sec.~\ref{sec:info_geo}, and Sec.~\ref{sec:management}).
\item Full three-dimensional sequential elasto-plastic analyses were operated in synchronization with excavation and instrumentation-installation steps, enabling consistent comparison between predicted and measured incremental responses (Sec.~\ref{sec:design}).
\item This synchronized analysis-and-measurement framework ensured traceability from initial assumptions to observations and to design revisions, while maintaining safety and optimizing schedule and cost under evolving ground conditions (Sec.~\ref{sec:issues}).
\end{itemize}

\subsubsection*{(ii) New insights and lessons learned} 
\noindent
Beyond the established methodology, the project yielded new technical insights that were not fully anticipated at the design stage:

\begin{itemize}
\item Behavior of hard rock governed by discontinuities: Even in generally competent hard rock, cavern-scale behavior was strongly influenced by major weak layers and discontinuities. Limitations of treating the rock mass as a homogeneous continuum were observed, requiring practical measures such as the introduction of interface elements and reconsideration of the relationship between averaged rock-mass deformability and strength (Sec.~\ref{sec:dome_design} and Sec.~\ref{sec:barrel_design}).
\item Integrity of support systems at large scale: At this cavern scale, maintaining the integrity of support elements, particularly prestressed anchors and shotcrete, was a key design and operational consideration (Sec.~\ref{sec:overload} and Sec.~\ref{sec:crack}).
\item Time-dependent behavior after excavation: Continuous post-blasting and post-excavation monitoring recorded time-dependent (creep) displacements, predominantly concentrated along major weak layers. The observed displacement rates decreased after excavation completion and showed convergent behavior (Sec.~\ref{sec:crack} and Sec.~\ref{sec:creep}).
\end{itemize}

Together, these observations summarize practical experience obtained from the excavation and monitoring of the Hyper-Kamiokande cavern and provide reference information for future large underground cavern projects under similar conditions, including potential scope for further examination of modeling approaches.

This cavern constitutes the fundamental part of the Hyper-Kamiokande experiment that aims to elucidate the history of cosmic evolution since the origin of the universe and to advance unified theories of elementary particles, serving as an international core facility for particle and astroparticle physics for more than two decades. 

\section*{Acknowledgments}
We thank Ministry of Education, Culture, Sports, Science and Technology and the University of Tokyo for financial support. We are grateful to the Hyper-Kamiokande collaborators for their commitment to the physics program and for their interest and support of the cavern works. 
We thank Eduardo de la Fuente Acosta, Katsuki Hiraide, Yoshitaka Itow, Luis Labarga, and Taku Ishida for constructive comments during the internal review and for careful content checks that strengthened this manuscript.
We acknowledge  Hyper-Kamiokande Project Advisory Committee under the directors of the Institute for Cosmic Ray Research 
(the University of Tokyo)  
and the Institute of Particle and Nuclear Studies (KEK), and in particular the Cavern Tank Subcommittee, for expert advice and guidance that were invaluable to safe completion. We also thank the Facilities Department of the University of Tokyo for their efforts in executing this unprecedented large-scale civil engineering project.
\let\doi\relax

\clearpage
\appendix
\section{ 
Supplementary schematics for excavation and support installation
}
This appendix provides supplementary schematics that support the description of the excavation and support installation. The figures and text in this appendix are referenced 
in 
the main text.

\begin{figure}[bthp!]
\begin{center}
\includegraphics[width=0.9\linewidth]{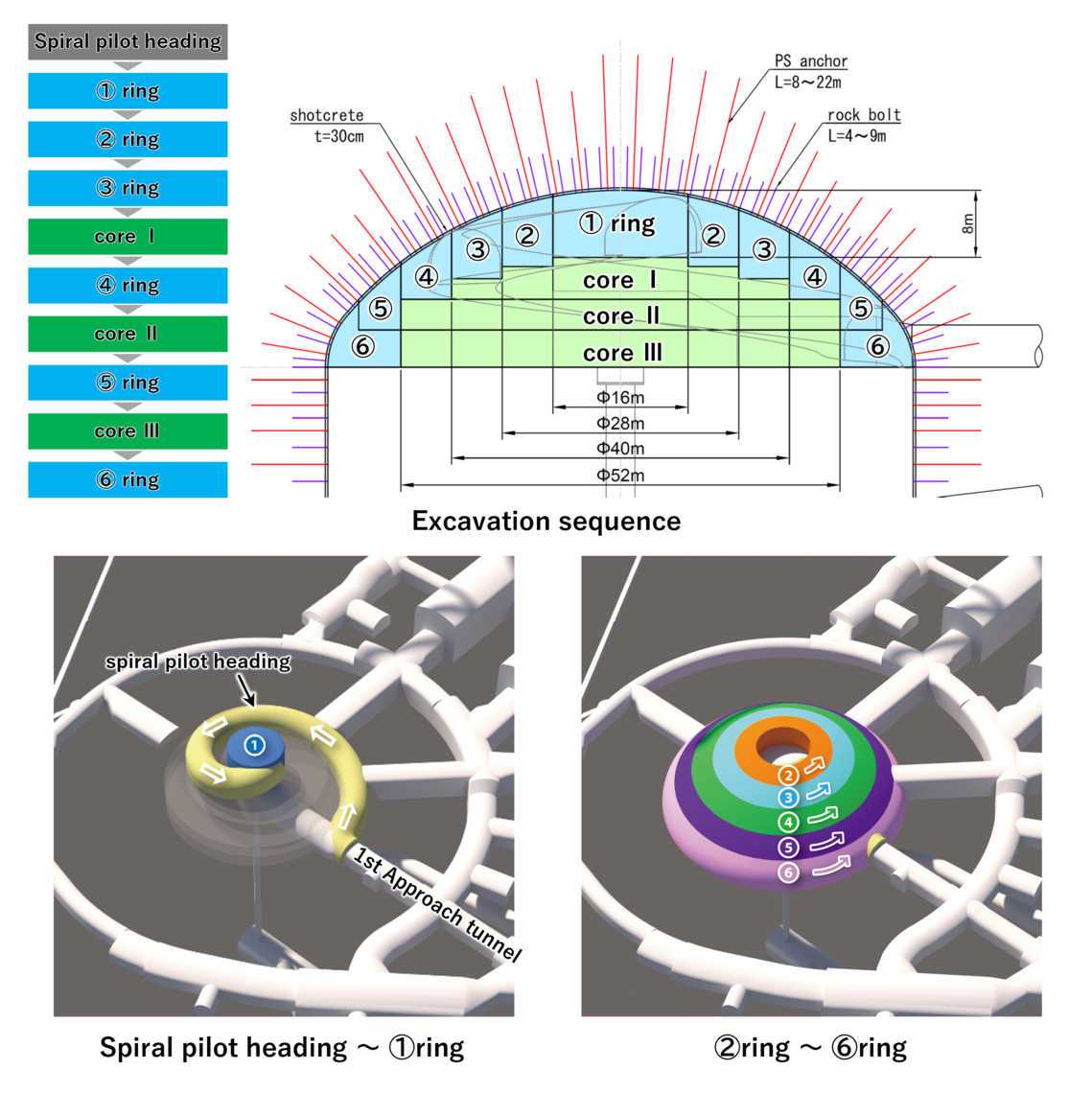}
\caption{
Schematic sequence of excavation and support installation in the dome section.
}
\label{fig:app_excavation_procedure}
\end{center}
\end{figure}
\clearpage 

\begin{figure}[bthp!]
\begin{center}
\includegraphics[width=1.0\linewidth]{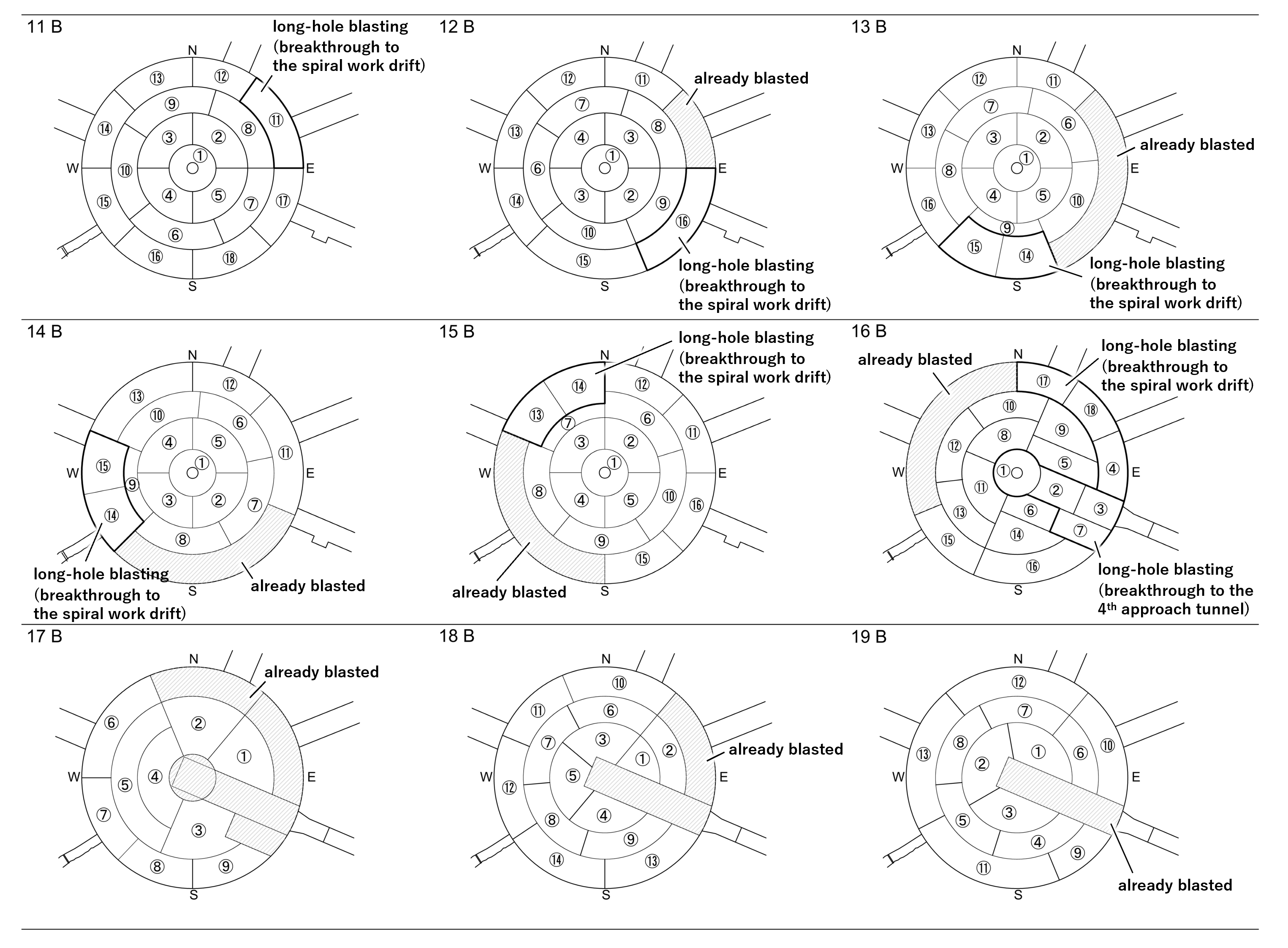}
\caption{
The as-built bench partitioning from 11B to 19B. 
Long-hole blasting blocks and 
already blasted blocks 
are also indicated in each diagram.
}
\label{fig:app_barrel_division}
\end{center}
\end{figure}

\subsection{
Excavation status at the onset of shotcrete cracking
} \label{sec:app_shotcrete_cracks}
Fig.~\ref{fig:app_barrel_division} provides the as-built bench partitioning for the cylindrical section and the block numbering used for long-hole blasting and bench blasting. At the time when the perimeter excavation of 12B was completed by long-hole blasting, shotcrete cracking was observed at the outermost dome ring (6R). Subsequently, after completion of blasting up to Block \textcircled{\footnotesize 9} (in 13B), additional cracking was identified along Weak Layer X7 between benches 4B and 6B, while countermeasures, support operations at the 12B wall, and the core excavation at 13B were carried out in parallel. Excavation was temporarily suspended for review and countermeasure planning.

To implement the countermeasures, restoration of the spiral working drift was required to re-establish access for equipment and operations. Therefore, the final core blasting at 13B (Block \textcircled{\footnotesize 10}) was executed, and the previously blasted sector (hatched area from the northeast to southeast in the 13B perimeter) was brought to completion through mucking, scaling, shotcreting, and installation of rock bolts and PS anchors.

\begin{figure}[bthp!]
\begin{center}
\includegraphics[width=1.0\linewidth]{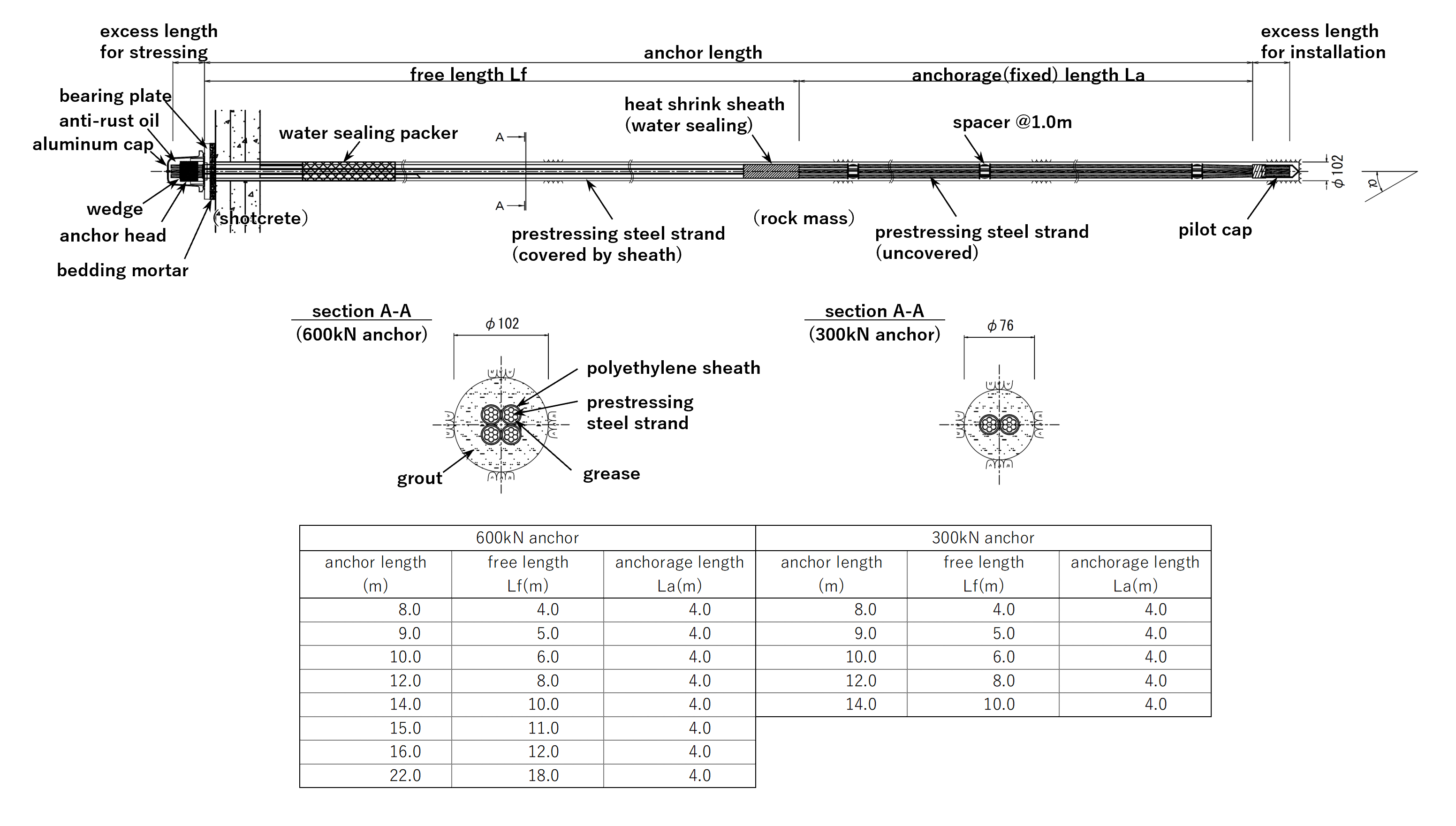}
\caption{
Structural details of the PS anchor (components, free/fixed lengths, and head assembly).
}
\label{fig:app_ps_anchor_structure}
\end{center}
\end{figure}

\begin{figure}[bthp!]
\begin{center}
\includegraphics[width=1.0\linewidth]{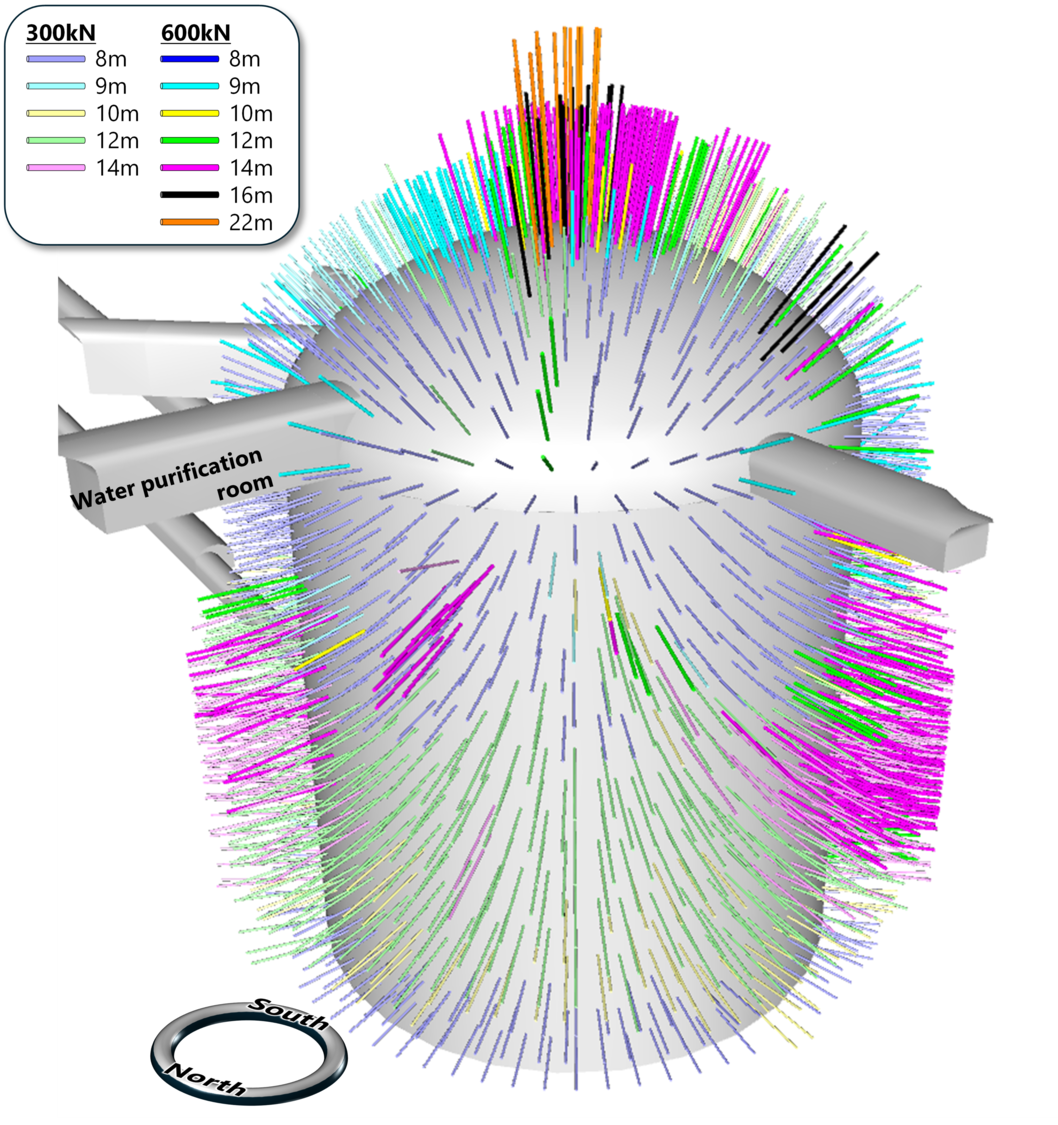}
\caption{
Bird's-eye view of the cavern illustrating the layout, orientations, and depths of the PS anchors, viewed from the northwest.
PS-anchor specifications (600 kN and 300 kN) and length categories are indicated in the legend and distinguished by color.
}
\label{fig:app_bird_ps_1}
\end{center}
\end{figure}

\begin{figure}[bthp!]
\begin{center}
\includegraphics[width=1.0\linewidth]{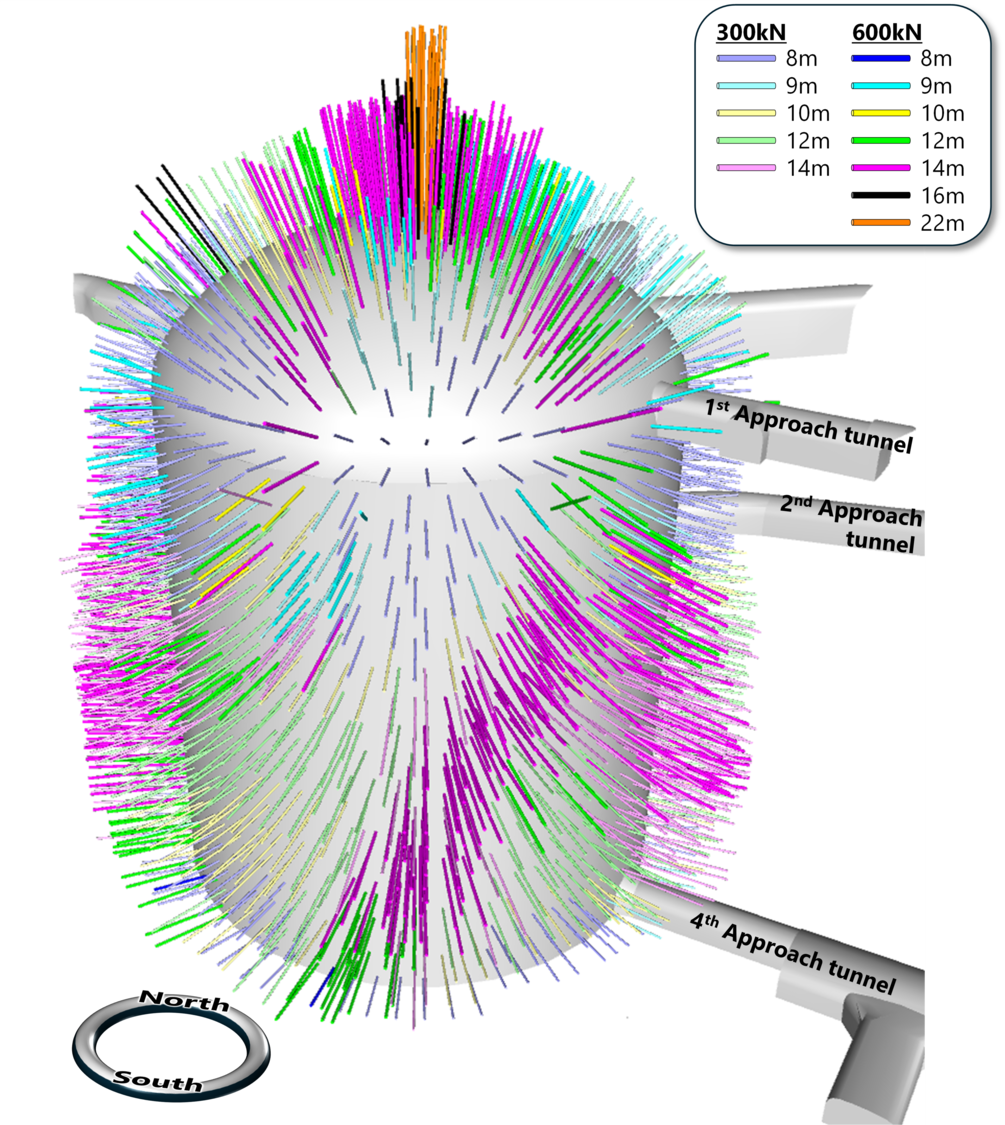}
\caption{
Bird's-eye view of the cavern illustrating the layout, orientations, and depths of the PS anchors, viewed from the southeast.
PS-anchor specifications (600 kN and 300 kN) and length categories are indicated in the legend and distinguished by color.
}
\label{fig:app_bird_ps_2}
\end{center}
\end{figure}

\clearpage 

\section{ 
Supplementary geological information
}
This appendix provides supplementary material that supports the description of the geological setting and data integration. The figures, table, and text in this appendix are referenced 
in 
the main text.

\begin{figure}[bthp!]
\begin{center}
\includegraphics[width=0.85\linewidth]{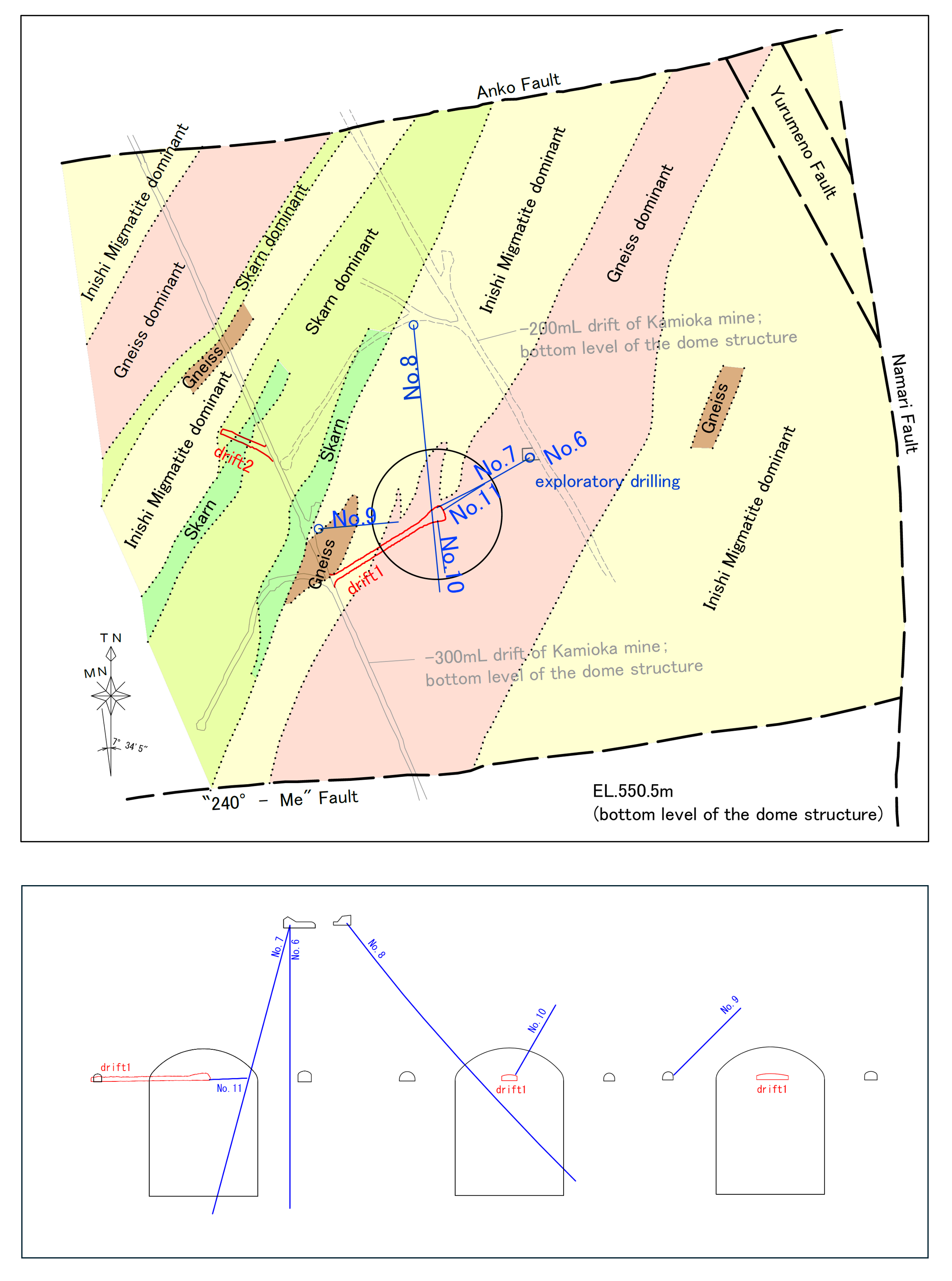}
\caption{
Regional geological map around the HK cavern and the relative locations of boreholes/adits and the cavern.
}
\label{fig:app_wide_geo_map}
\end{center}
\end{figure}

\begin{figure}[bthp!]
\begin{center}
\includegraphics[width=1.0\linewidth]{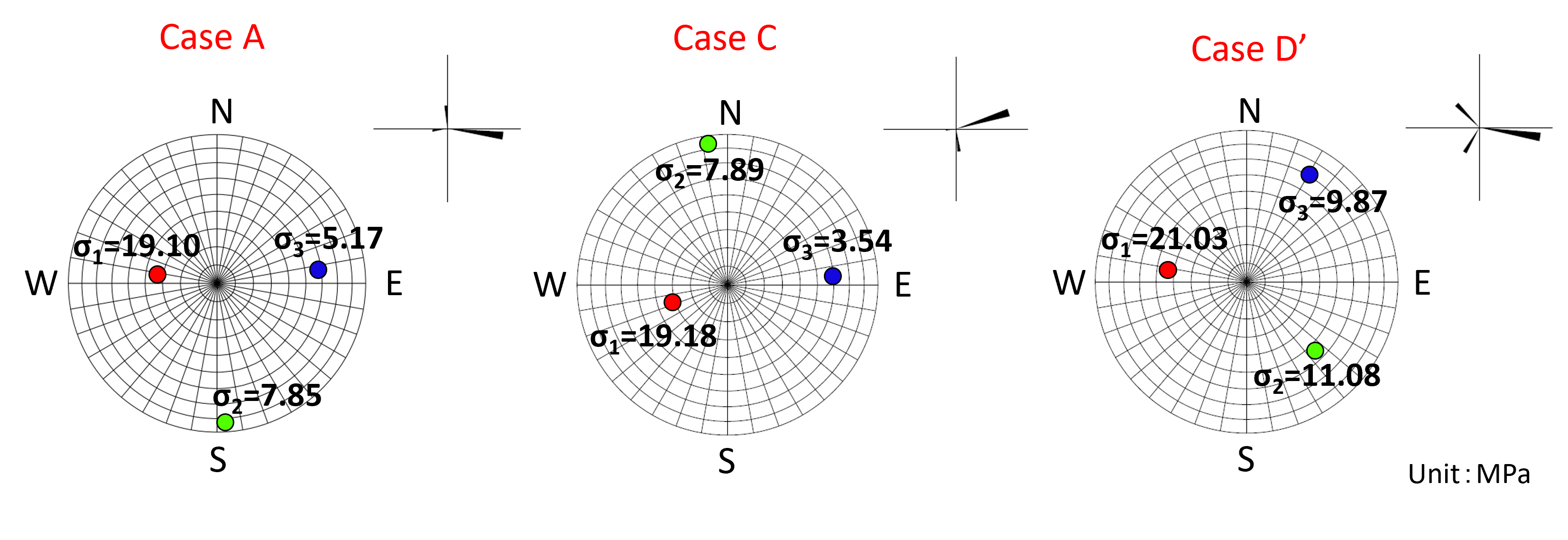}
\caption{
Principal stress orientations and magnitudes for the representative stress cases (Cases A, C, and D'). Each stereonet plot shows the directions of the major, intermediate, and minor principal stresses ($\sigma_1$, $\sigma_2$, $\sigma_3$) and their magnitudes (unit: MPa).
}
\label{fig:app_inist_dir}
\end{center}
\end{figure}

\begin{table}[htbp]
\centering
\caption{
Definition of rock-mass classes (based on \cite{rockclass}).
}
\label{tab:app_rock_mass_classification}
\begin{small}
\hspace*{-0.5cm}
\begin{tabular}{|c|p{3.5cm}|p{2.0cm}|p{3.0cm}|p{4.5cm}|}
\hline
\textbf{\makecell{Rock-mass \\ class}} 
&
\textbf{Hardness / Rock condition} 
&
\textbf{Joint spacing} 
&
\textbf{Joint condition} 
&
\textbf{Borehole core characteristics} 
\\
\hline
B
 &
Fresh rock; metallic sound when struck by hammer (very hard and dense). Constituent minerals are not weathered or altered. 
&
Sparse 
&
Tight 
&
Core length typically around 40--50~cm; rock is fresh, joints are absent or tightly closed; core recovery is excellent. 
\\
\hline
CH 
&
Generally fresh rock; metallic sound when struck by hammer (hard). Constituent minerals may show slight weathering or alteration. 
&
Many joints 
&
Generally tight (joint surfaces may be discolored due to weathering or alteration) 
&
Core length typically around 10--30~cm; rock is generally fresh; joints are moderately developed; joint surfaces are often light brown due to weathering, but alteration does not extend into the core interior; core recovery is good. 
\\
\hline
CM 
&
Moderately weathered or altered rock; often breaks along joints under strong hammer blows. Constituent minerals are weathered and show brownish discoloration. 
&
Numerous closely spaced joints 
&
Open (often with clay infill) 
&
Core length typically around 10~cm; rock is moderately weathered; joint surfaces are weathered and alteration extends into the interior; core recovery is typically greater than 80\%. 
\\
\hline
CL 
&
Strongly weathered and altered rock; brownish coloration; easily disintegrates under light hammer blows. 
&
Very dense distribution of small joints 
&
Open (strong brown discoloration and abundant clay infill along joints) 
&
Fragmented core; rock is strongly weathered throughout; core recovery is typically less than 80\%. 
\\
\hline
\end{tabular}
\end{small}
\end{table}

\subsection{
GSI framework
} \label{sec:app_gsi}
The GSI (Geological Strength Index) system is a rock-mass property evaluation framework used to estimate the peak strength and elastic modulus of a rock mass based on rock core characteristics and geological conditions, including jointing features. These rock-mass properties are estimated by combining results of laboratory rock core tests with geological observations obtained from boreholes and excavated surfaces.

\begin{figure}[bthp!]
\begin{center}
\includegraphics[width=1.0\linewidth]{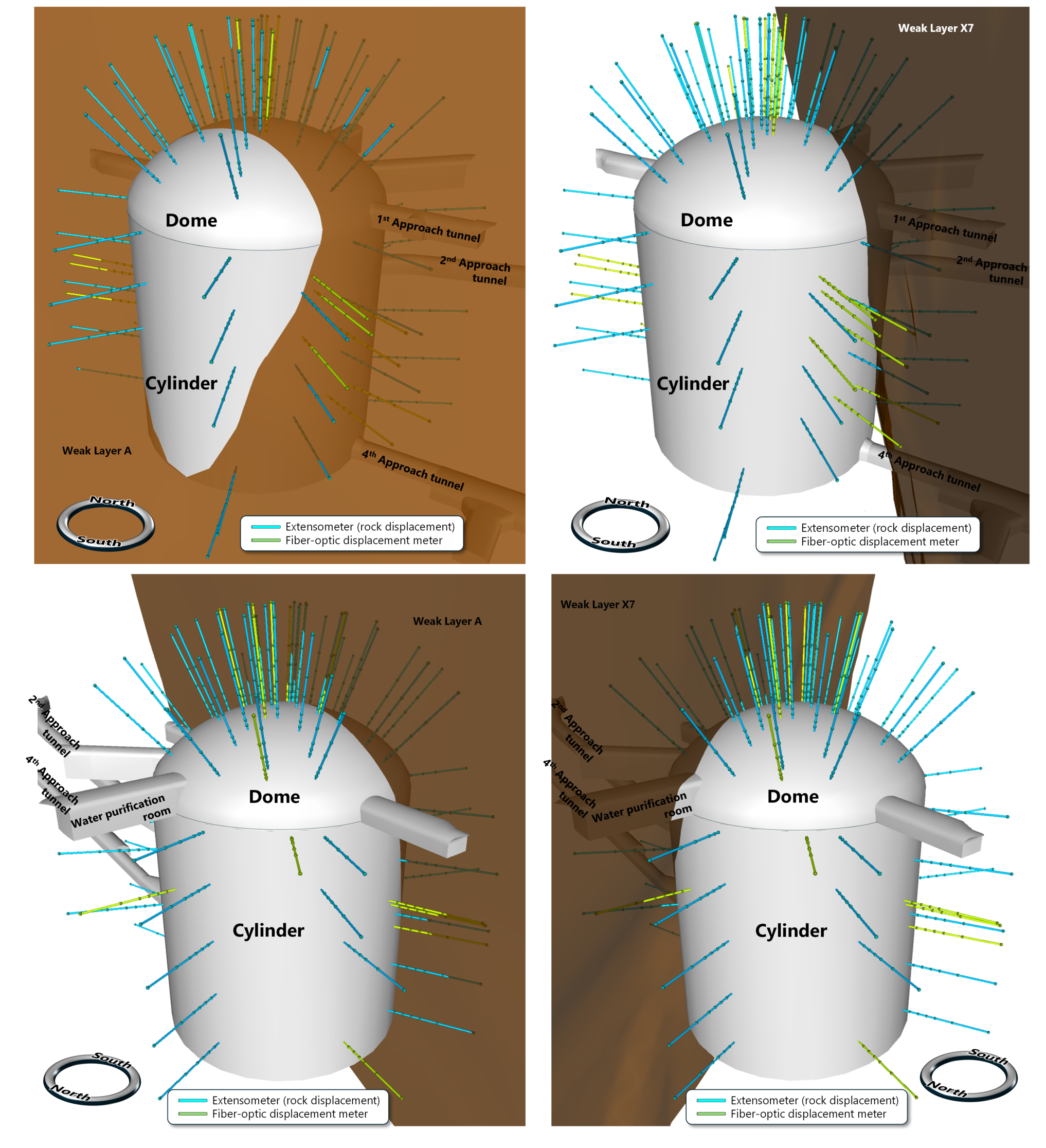}
\caption{
Bird's-eye views of the cavern illustrating the relationship with the major weak layers.
The left panels show Weak Layer A, and the right panels show Weak Layer X7. The top panels are viewed from the southeast, and the bottom panels are viewed from the northwest.
All panels also illustrate the locations, orientations, and depths of the extensometers (cyan) and fiber-optic displacement meters (green). Knots on the extensometers indicate the fixed points used for displacement measurements.
}
\label{fig:app_bird_inst_wl}
\end{center}
\end{figure}

\clearpage
\section{ 
Conceptual diagrams for support design
}
This appendix provides supplementary schematics that support the description of the support design concept and stability-check framework. The figures and text in this appendix are referenced
in 
the main text.

\begin{table}[htbp]
\centering
\caption{
Schematic illustrations of the design concepts 
to identify potentially unstable rock blocks and wedges considered in the stability checks~\cite{ARMS14}.
}
\label{tab:app_sup_design_dome}
\begin{small}
\begin{tabular}{p{5cm}p{0.1cm}p{5.0cm}p{0.1cm}p{5.0cm}}
\hline
Suspension capacity
& &
Prevention of dislodgement
& &
Tensile-failure block collapse
\\
\hline
\begin{minipage}{0.3\textwidth}
\centering
\includegraphics[width=1.0\linewidth]{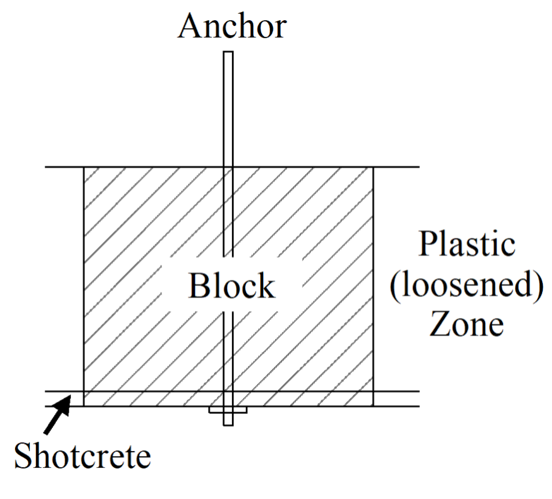}
\end{minipage}
& &
\begin{minipage}{0.3\textwidth}
\centering
\includegraphics[width=1.0\linewidth]{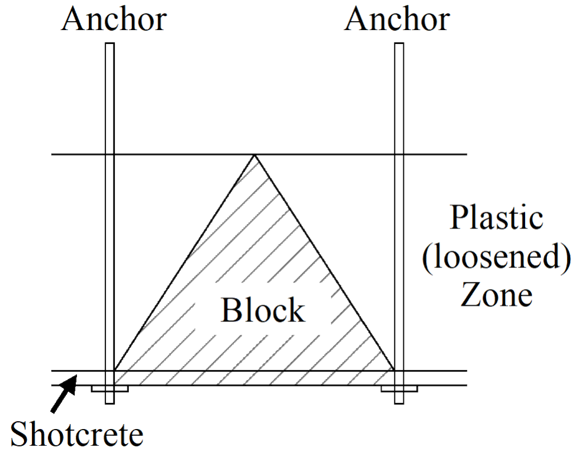}
\end{minipage}
& &
\begin{minipage}{0.3\textwidth}
\centering
\includegraphics[width=0.9\linewidth]{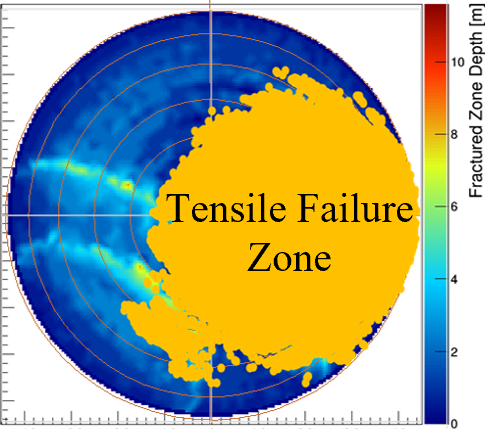}
\end{minipage}
\\
\hline
Each PS anchor must support the weight of the rock block, using pretension and shotcrete shear resistance.
& &
Rock between anchors must be safely supported by shotcrete shear capacity.
& &
The block is modeled as a uniform-thickness block extending through the depth of the plastic (loosened) zone.
\\
\hline
\\
\hline
& &
Shear-failure wedge collapse
\\
\hline
\multicolumn{1}{m{5.0cm}}{
\begin{minipage}[t]{0.3\textwidth}
\centering
\vspace*{0.7cm}
\includegraphics[width=1.1\linewidth]{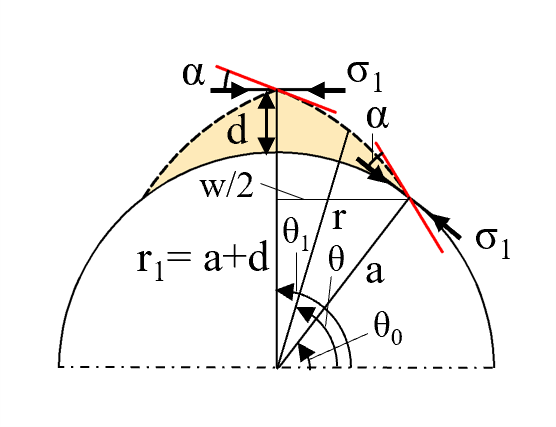}
\end{minipage}
}
& &
\multicolumn{1}{p{5.0cm}}{\footnotesize
\begin{minipage}[t]{0.3\textwidth}
Parameter relationship:
\[ 
\begin{array}{ll}
(1) & \displaystyle \alpha = \frac{\pi}{4} - \frac{\Phi}{2}, \\
(2) & \displaystyle r = a \cdot e^{(\theta - \theta_0)\tan\alpha}.
\end{array}
\]
Here, $\theta$ is the polar angle measured from the horizontal axis, and $r$ is the distance from the cavern center to the mobilized joint plane; $w$ can be obtained as follows:
\[
\begin{array}{ll}
(3) & \displaystyle r_1 = a + d = a \cdot e^{(\theta_1 - \theta_0)\tan\alpha}, \\
(4) & \displaystyle w = 2 a \sin (\theta_1 - \theta_0).
\end{array}
\]
\end{minipage}
}
& & 
\multicolumn{1}{p{5.0cm}}{\footnotesize
\begin{minipage}[t]{0.3\textwidth}
Definition of parameters:
\begin{description}
\item [$\alpha$: ] the angle between the mobilized joint plane and the $\sigma_1$ direction,
\item [$a$: ] cavern radius,
\item [$\theta_0$: ] polar angle at $r=a$,
\item [$d$: ] loosened zone depth at the ridge no the mobilized joint plane,
\item [$r_1$: ] distance from the cavern center to the ridge,
\item [$\theta_1$: ] polar angle to the ridge,
\item [$w$: ] circumferential wedge-block width (chord length at radius $a$).
\end{description}
\end{minipage}
}
\\
\hline
\multicolumn{5}{p{17.2cm}}{
For shear zones due to circumferential uniaxialization, stability was verified for wedges formed along mobilized joint planes determined by friction angle $\Phi$. 
Using a logarithmic (equiangular) spiral, the circumferential wedge width ($w$) is defined as described above.
}
\\
\hline
\end{tabular}

\end{small}
\end{table}

\begin{figure}[bthp!]
\begin{center}
\includegraphics[width=1.0\linewidth]{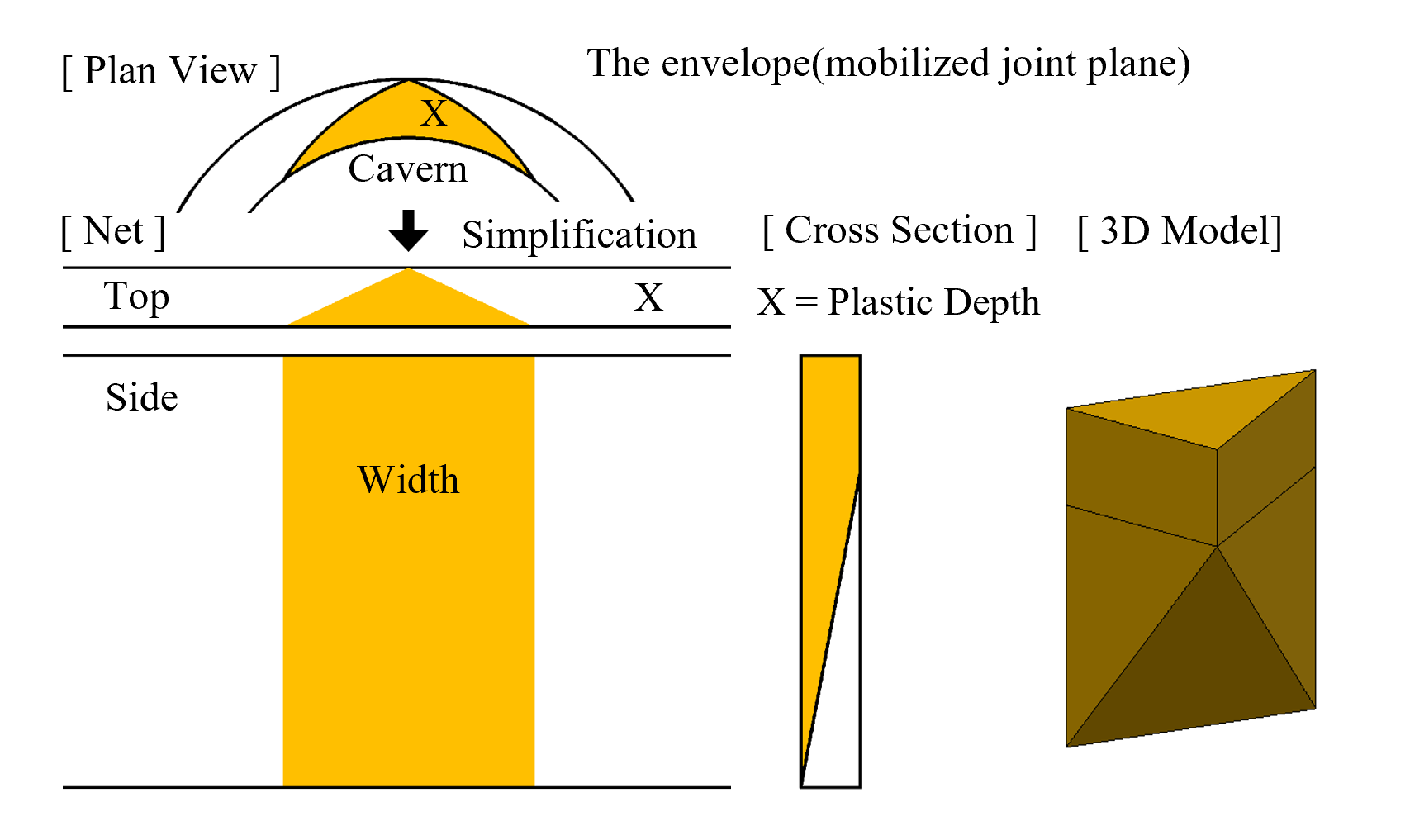}
\caption{
Schematic illustration of the simplified 3D sliding-wedge block model on the cylindrical wall~\cite{ARMS14}.
The circumferential wedge width is defined using Eqs. (1)--(4) in Table ~\ref{tab:app_sup_design_dome}; in earlier 2D practice, no explicit wedge width is defined and a constant-thickness sliding block is assumed. The 3D definition allows shear resistance on the vertical shotcrete faces to be included.
}
\label{fig:app_3d_block}
\end{center}
\end{figure}

\clearpage
\section{ 
Instrumentation details and layout
}
This appendix provides supplementary material that supports the description of the monitoring instruments and their layout. The figures and text in this appendix are referenced
in 
the main text.

\begin{figure}[bthp!]
\begin{center}
\includegraphics[width=1.0\linewidth]{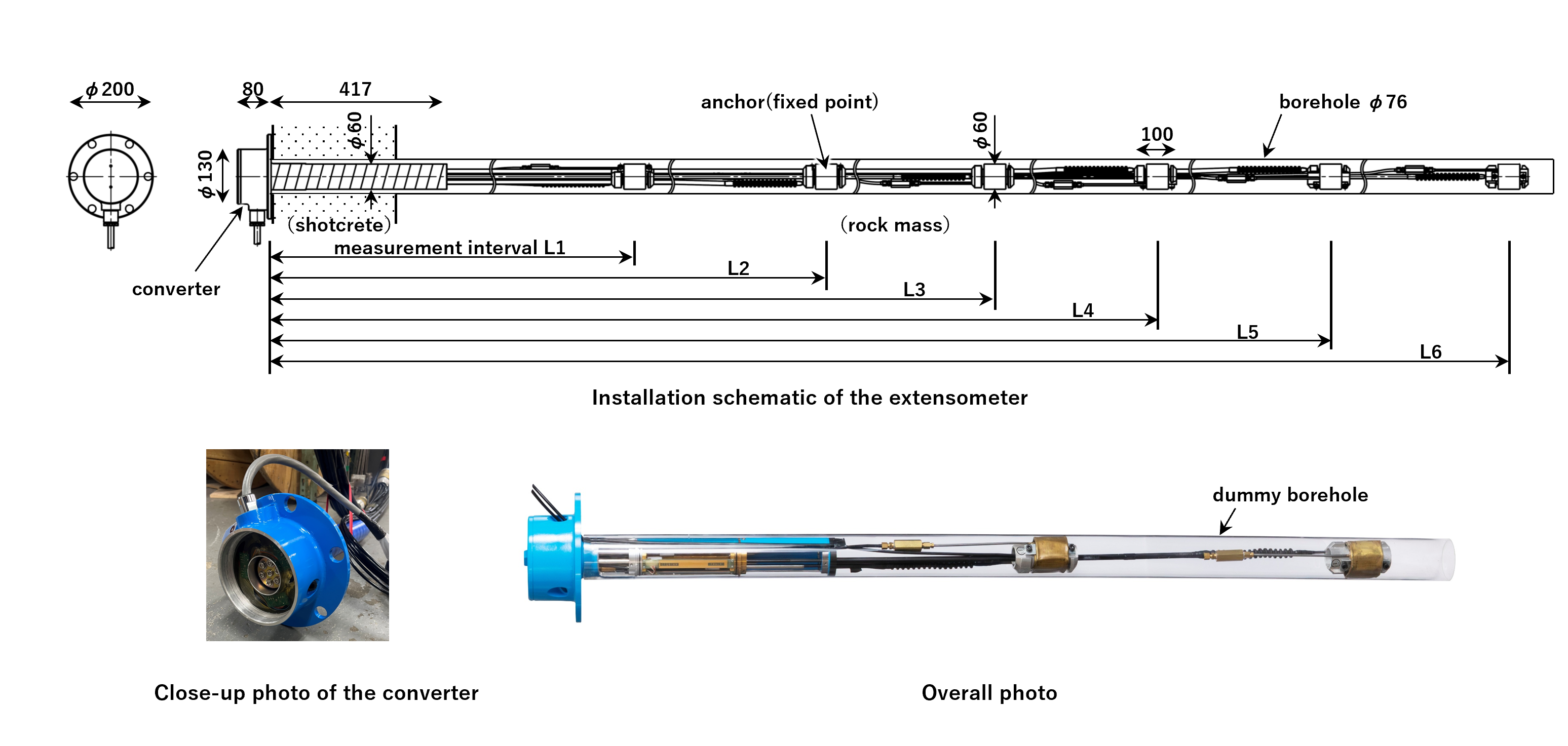}
\caption{
Schematic diagrams and photographs of the extensometer.
}
\label{fig:app_extensometer}
\end{center}
\end{figure}

\begin{figure}[bthp!]
\begin{center}
\includegraphics[width=1.0\linewidth]{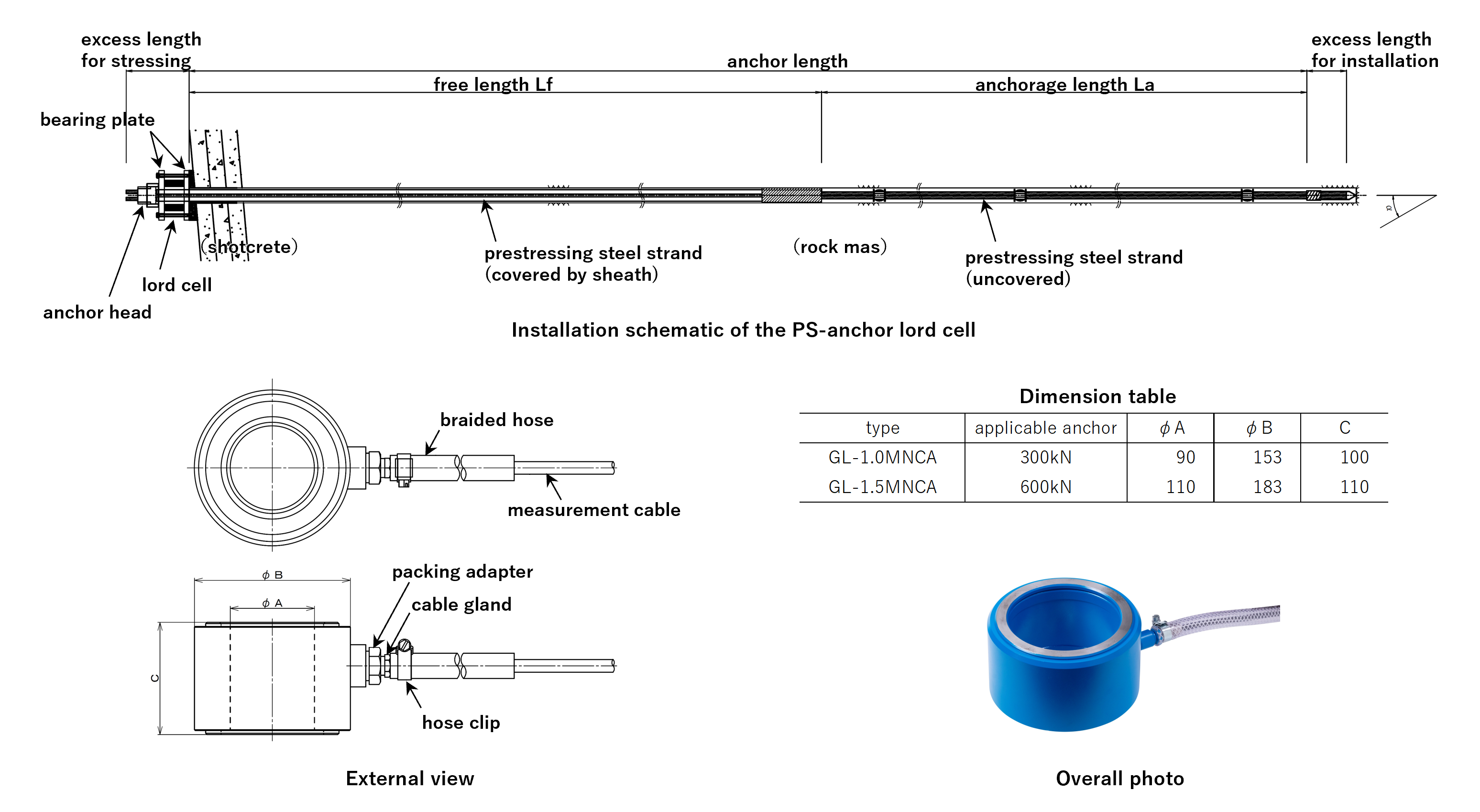}
\caption{
Schematic diagrams and a photograph of the PS-anchor load cell.
}
\label{fig:app_load_cell}
\end{center}
\end{figure}

\begin{figure}[bthp!]
\begin{center}
\includegraphics[width=1.0\linewidth]{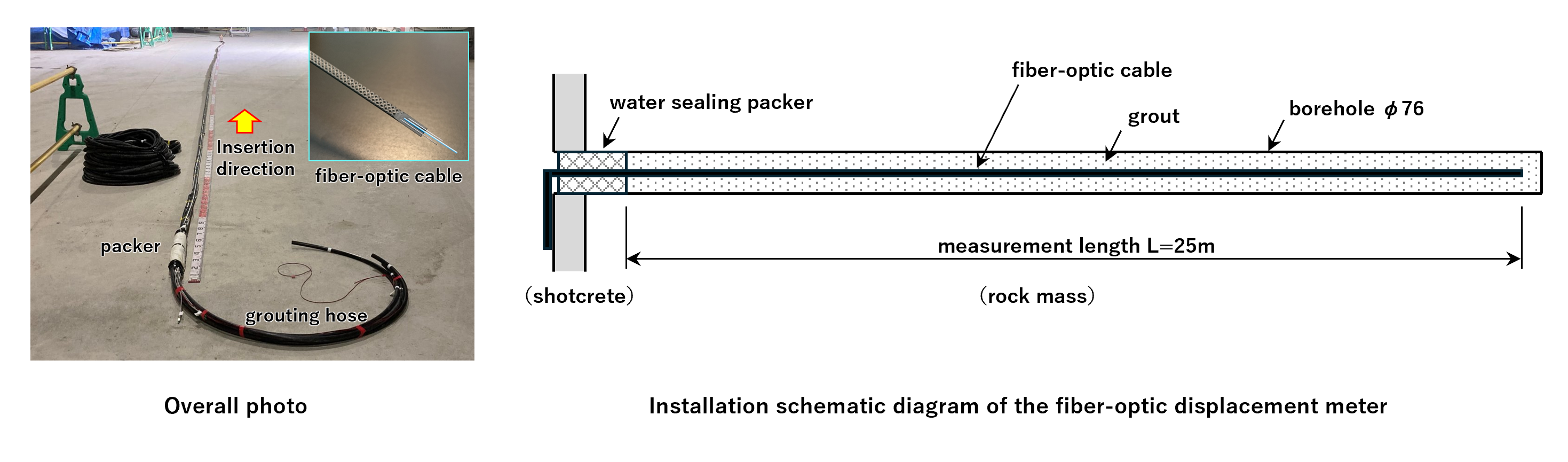}
\caption{
Schematic diagrams and a photograph of the fiber-optic displacement meters~\cite{fiber_displacement_sensor}.
}
\label{fig:app_fiber_sensor}
\end{center}
\end{figure}

\begin{figure}[bthp!]
\begin{center}
\includegraphics[width=1.0\linewidth]{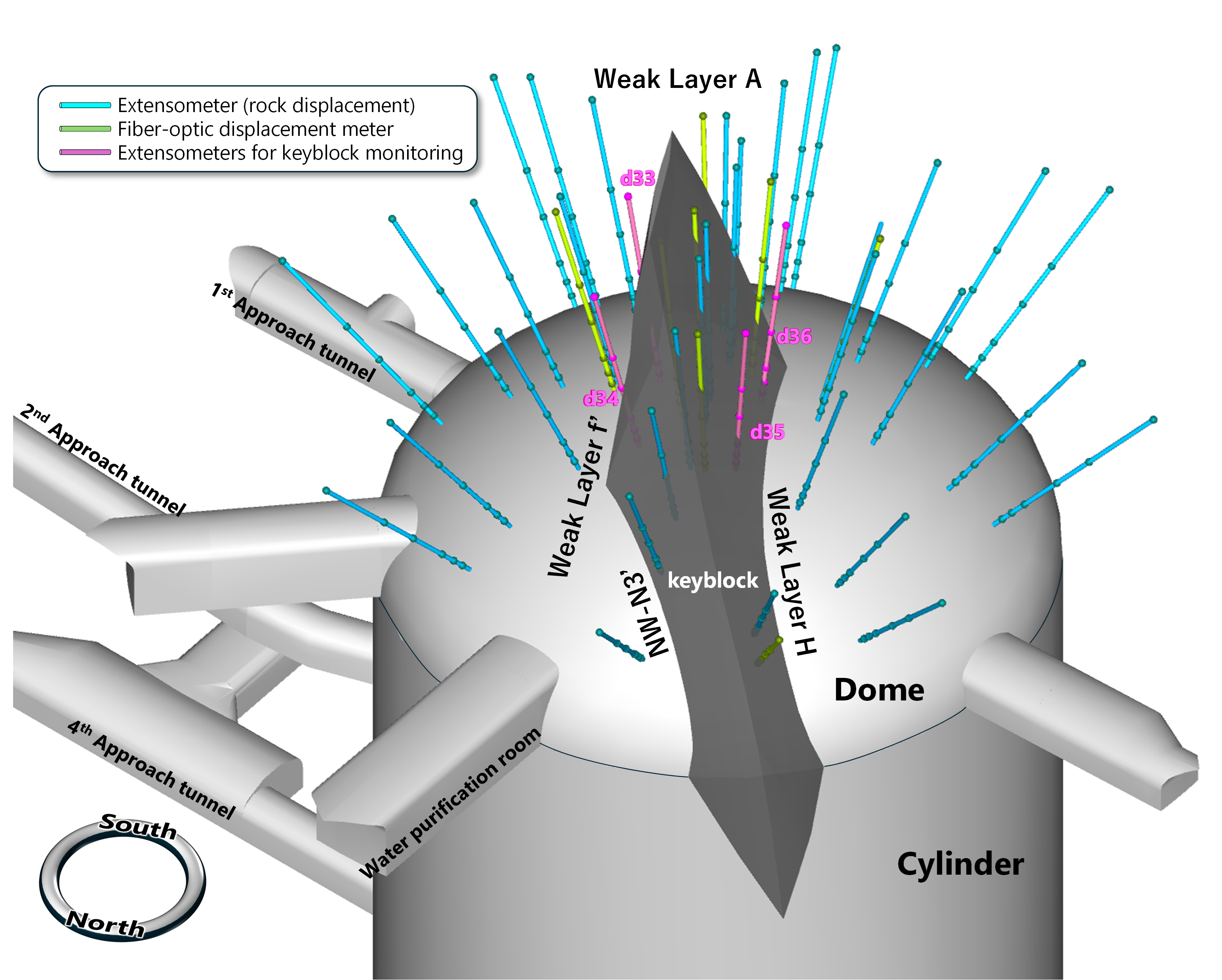}
\caption{
Bird's-eye view of the dome section illustrating the relationship between the assumed massive key block (open on the northeast side) and extensometers d33--d36.
Knots on the extensometers indicate the fixed points used for displacement measurements.
}
\label{fig:app_dome_keyblock}
\end{center}
\end{figure}

\end{document}